\documentclass[11pt]{article}
\usepackage[left=1in,top=1in,right=1in,bottom=1in,head=.1in,nofoot]{geometry}

\usepackage{setspace,url,bm,amsmath} 
\usepackage{float}
\usepackage{caption}
\usepackage{subcaption,siunitx,booktabs}
\usepackage{enumerate}

\usepackage{titlesec} 
\usepackage{rotating}
\titlelabel{\thetitle.\quad} 
\titleformat*{\section}{\bf\large}

\usepackage{graphicx} 
\usepackage{bbm}
\usepackage{latexsym}
\usepackage{color}
\usepackage{authblk}
\usepackage{amsthm}
\usepackage{amsfonts}

\usepackage{amsthm}
\usepackage{amsfonts}
\theoremstyle{definition}
\newtheorem{proposition}{Proposition}

\newtheorem{corollary}{Corollary}

\usepackage{natbib} 
\bibpunct{(}{)}{;}{a}{}{,} 
\usepackage{enumitem}

\usepackage{etoolbox} 
\apptocmd{\sloppy}{\hbadness 10000\relax}{}{} 

\def\T{\textrm{T}}
\def\new{\textrm{new}}

\begin{document}

\doublespacing
\frenchspacing
\title{\bf An Email Experiment to Identify the Effect of Racial Discrimination on Access to Lawyers: A Statistical Approach}

\author[*]{Brian Libgober}
\author[**]{Tirthankar Dasgupta}

\affil[*]{Department of Political Science, Yale University, 115 Prospect Street, Rosenkranz Hall, New Haven, CT 06520-8301, email: brian.libgober@yale.edu}
\affil[***]{Department of Statistics, Rutgers University, 110 Frelinghuysen Road, Piscataway, NJ 08901, email: tirthankar.dasgupta@rutgers.edu}

\renewcommand\Authands{ and }

\date{}
\maketitle

\begin{abstract}
\frenchspacing
We consider the problem of conducting an experiment to study the prevalence of racial bias against individuals seeking legal assistance, in particular whether lawyers use clues about a potential client's race in deciding whether to reply to e-mail requests for representations. The problem of discriminating between potential linear and non-linear effects of a racial signal is formulated as a statistical inference problem, whose objective is to infer a parameter determining the shape of a specific function. Various complexities associated with the design and analysis of this experiment are handled by applying a novel combination of rigorous, semi-rigorous and rudimentary statistical techniques. The actual experiment was attempted with a population of lawyers in Florida, but could not be performed with the desired sample size due to resource limitations. Nonetheless, it provides a nice demonstration of the proposed steps involved in conducting such a study. 
\end{abstract}


\section{Introduction}
\frenchspacing

It is widely thought in the legal community that there is a serious ``access gap'' along racial and economic lines. Poorer individuals and minorities have a harder time resolving their legal problems for many reasons, including the high cost of legal services, the historic under-representation of minorities in the legal profession, and the incomprehensibility and hostility of the legal system to outsiders. Lack of access to the legal system likely contributes to predatory behavior by malignant actors, so that problems of some groups in securing representation are plausible contributing factors to diverse social ills such as mass incarceration, predatory lending, workplace discrimination, police brutality, and others. What is unclear, however, is whether the problems that minorities have in accessing lawyers are purely a consequence of lack of resources, or whether the mostly white legal profession discriminates against potential clients using race.

Whether minorities face discrimination in access to lawyers is best established through carefully designed experiments. E-mail experiments have been used to investigate racial access barriers in many contexts, from employment \citep{Emily2004}, to housing \citep{Carpusor2006}, to voting \citep{White2015,Butler2011}, to getting advice on graduate school \citep{Milkman2015}, both in the American context and abroad \citep{Carlsson2007}. Excellent recent surveys of this literature include  \cite{Bertrand2017} and \cite{Gaddis2018a}. In contrast with non-experimental studies of racial discrimination, or experimental studies that use actors posing as genuine applicants \citep[e.g.][]{Pager2006a}, an important benefit of e-mail experiments is that they raise much less concern about the effect of unobservable confounders. Though such studies are still not immune from criticism about whether they measure quantities that are economically meaningful \citep[see, e.g.,][]{Heckman1998}, many consider them as offering some of the clearest evidence that, at least of the time, there can be substantively large differences in access by race that are persistent over time \citep{Quillian2017}.

To explore the feasibility and prospects of such an experiment, a small pilot study patterned after the resume studies popularized by \cite{Emily2004} was conducted by the authors of this paper. In this study, email addresses for lawyers were collected from the California bar association's websites. Messages were sent to the lawyer asking if he or she would consider taking a case. The treatment intervention in this experiment was the name of the requester, which was highly suggestive of the client's race. For example, one lawyer might have received an inquiry from Darnell Jackson, who felt he was wrongfully found guilty of driving under the influence. Another lawyer might have received an email that was exactly similar in content, but sent by an individual named Brad McCarthy.

The pilot study indicated a probable effect of a potential client's race on the probability of a lawyer responding to that client's request. The names used in the pilot experiment were taken from literature, and belonged to two groups. Names belonging to one group were assumed to be associated with black with certainty. Similarly, those belonging to the other group were assumed to be associated with white with certainty. Thus, in this study, the input factor race was essentially studied at two levels - 0 (black with certainty) and 1 (white with certainty).

A novel aspect of our research design in contrast to our pilot study and some previous email correspondence studies  is consideration of more ``levels'' of the treatment. Critiquing existing aduit studies, \cite{Gaddis2017} writes that typically ``researchers take a shortcut by first using a specific subset of names and then taking a continuous variable of racial naming practices and turning it into a binary (i.e., white name or black name).'' \cite{Emily2004}, for example, only used names that approximately thirty out of thirty people would say belong to a black person and names that almost thirty out of thirty people would say belong to a white person. However, most individuals do not have names that are racially identified to such a high degree. It is reasonable to suppose that a disparate racial effect must exist to a lesser, but still significant, degree for individuals with names
that are less racially identified. Yet exactly how clues about race translate into disparities in treatment remains speculative \citep{Gaddis2017a, Gaddis2017}. Perhaps only individuals whose names are ``very black'' or ``very white'' get treated differently because of that racial association, while the majority of people with
names that are less strongly associated do not seem to get treated much differently at all. Alternatively, perhaps even modest hints about race trigger vastly different behavior by service providers. Is the discriminatory impact identified by email experiments confined only
to the extreme levels typically studied? With an experiment only done at two extreme levels, one has no
power to confirm or refute such hypotheses. Although many studies \citep[such as][]{Carpusor2006,Carlsson2007} have shown that discrimination at two extreme levels does exist in many context, none to our knowledge has considered the more nuanced question about what outcomes follow when race is signaled to varying degrees.

Conducting such an experiment is clearly a non-trivial problem that involves several interesting and challenging questions related to its design and analysis. The design questions are: (i) how many levels of treatments do we need to consider? (ii) how many experimental units (lawyers) do we need to expose to each level of treatment? (iii) how does one physically administer a treatment level, that is, identify a name with an intended racial signal? (iv) how to account for the diversity among the lawyers in terms of type of practice, gender and race? From the analysis perspective, the most important questions are (i) what type of statistical model should one consider? and (ii) how to perform valid statistical inference from the experimental data in a manner that is consistent with the design? The first two design questions and the two analysis questions are inseparable in the sense that they depend on each other. To the best of our knowledge, no formal or rigorous statistical framework exists that collectively answers the above questions and consequently enables one to design such email experiments. 

In this article, we describe a comprehensive approach that addresses all the problems described above, and demonstrate how such an approach was used to plan the first experiment of its kind. Unfortunately, due to resource limitations, the plan could not be implemented with adequate sample size that would guarantee sufficient power to identify treatment effects. However, it may still be considered as a pioneering effort of its kind and provides a demonstration of the proposed approach. It is worthwhile to note that although this specific methodology was developed in the context of legal experiments, it is applicable to a broader class of social experiments involving assessment of bias on a service outcome.

In the following Section, we provide a statistical formulation of the problem. In Section \ref{Sec:MLE_Fisher}, we describe the statistical inference procedure assuming that we have experimental data, and in Section \ref{Sec:Design} address the four key design questions using theoretical results of Section \ref{Sec:MLE_Fisher} and extensive simulation studies. In Section \ref{s:analysis}, we demonstrate analysis of the experimental data. We conclude with a discussion in Section \ref{Sec:Discussion}. 

\section{Statistical Formulation} \label{Sec:stat_formulation}

Consider a set $\mathcal{R}$ of service-providers (henceforth referred to as the \textit{receiver}), each of whom is either black or white. Also, consider a set of names $\mathcal{S}$ associated with service-seekers (henceforth referred to as \textit{senders}), each of which can be associated with a black or a white person. Assume that each receiver in $\mathcal{R}$, irrespective of his/her own race, will associate a name in $S$ with a person of black origin with the same probability $\xi \in \left( 0,1\right) $ and with a person of white origin with probability $1- \xi$. We can therefore assign a number called the ``race level'' $\xi \in \left(0,1\right)$ to a specific name X in $\mathcal{S}$ which which a receiver will identify it as black. As an example, a name $X$ with race level $\xi = 0.6$ is one that can be identified as black by a receiver with probability 0.6. We further assume that in reality there does not exist a name with either $\xi=0$ or $\xi=1$, although there can exist names for which $\xi$ can be very close to 0 or 1. 

\begin{figure}[ht]
\centering
\caption{Plausible non-linear behaviors of increasing $\pi(\xi)$} \label{fig:two_functions1}
\begin{tabular}{cc}
\includegraphics[scale=0.45]{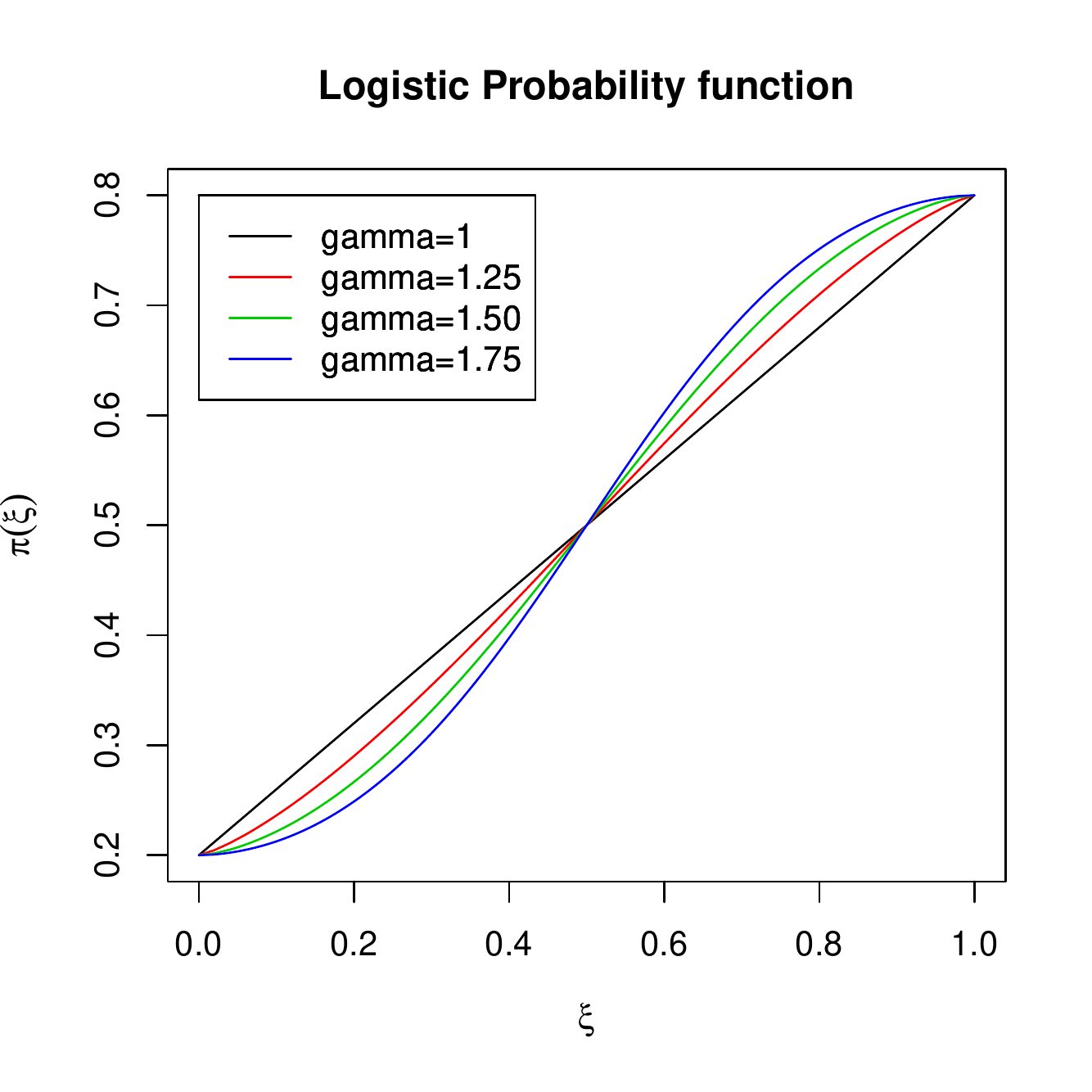} &
\includegraphics[scale=0.45]{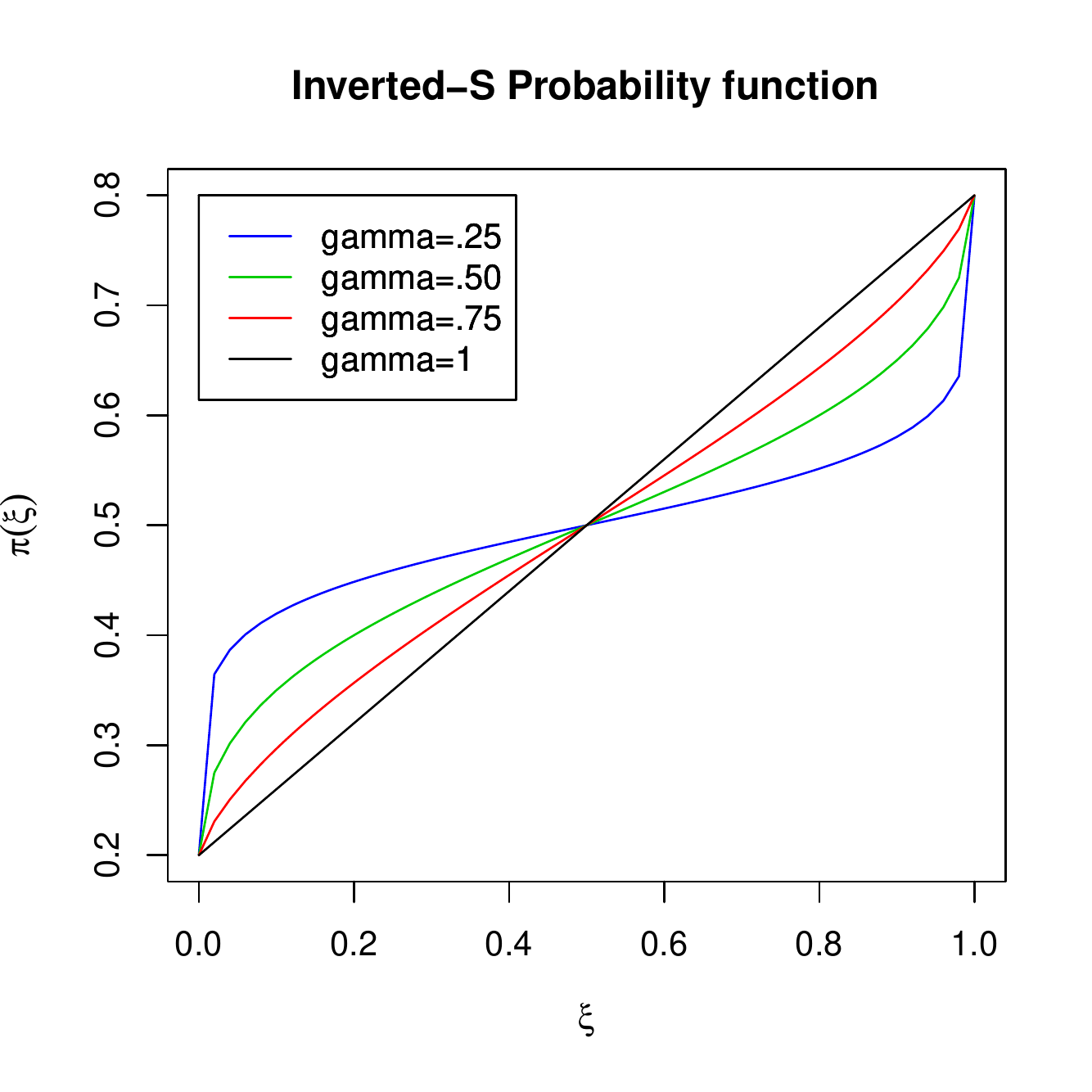} 
\end{tabular}
\end{figure}

\begin{figure}[ht]
\centering
\caption{Plausible non-linear behaviors of decreasing $\pi(\xi)$} \label{fig:two_functions2}
\begin{tabular}{cc}
\includegraphics[scale=0.45]{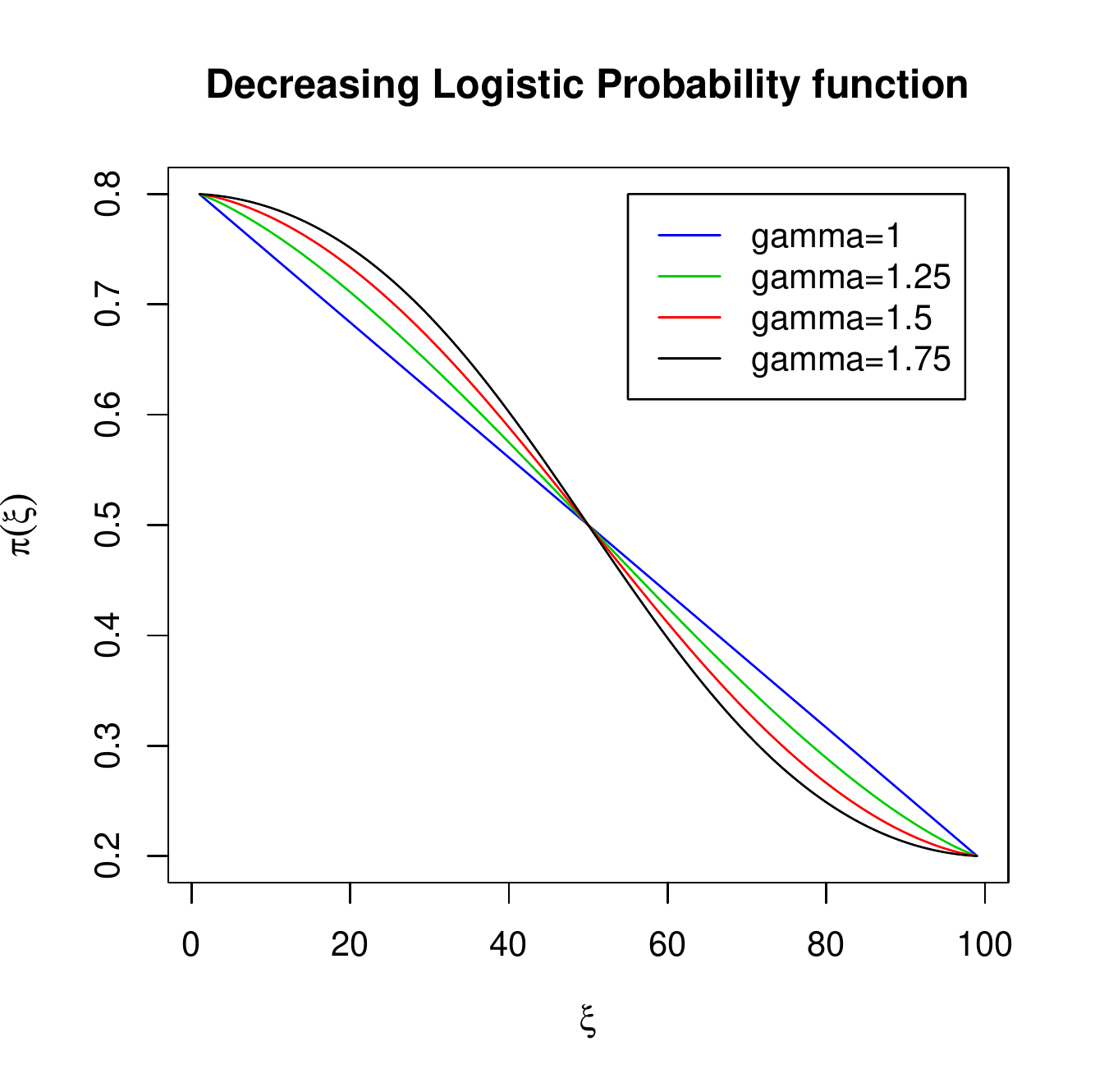} &
\includegraphics[scale=0.45]{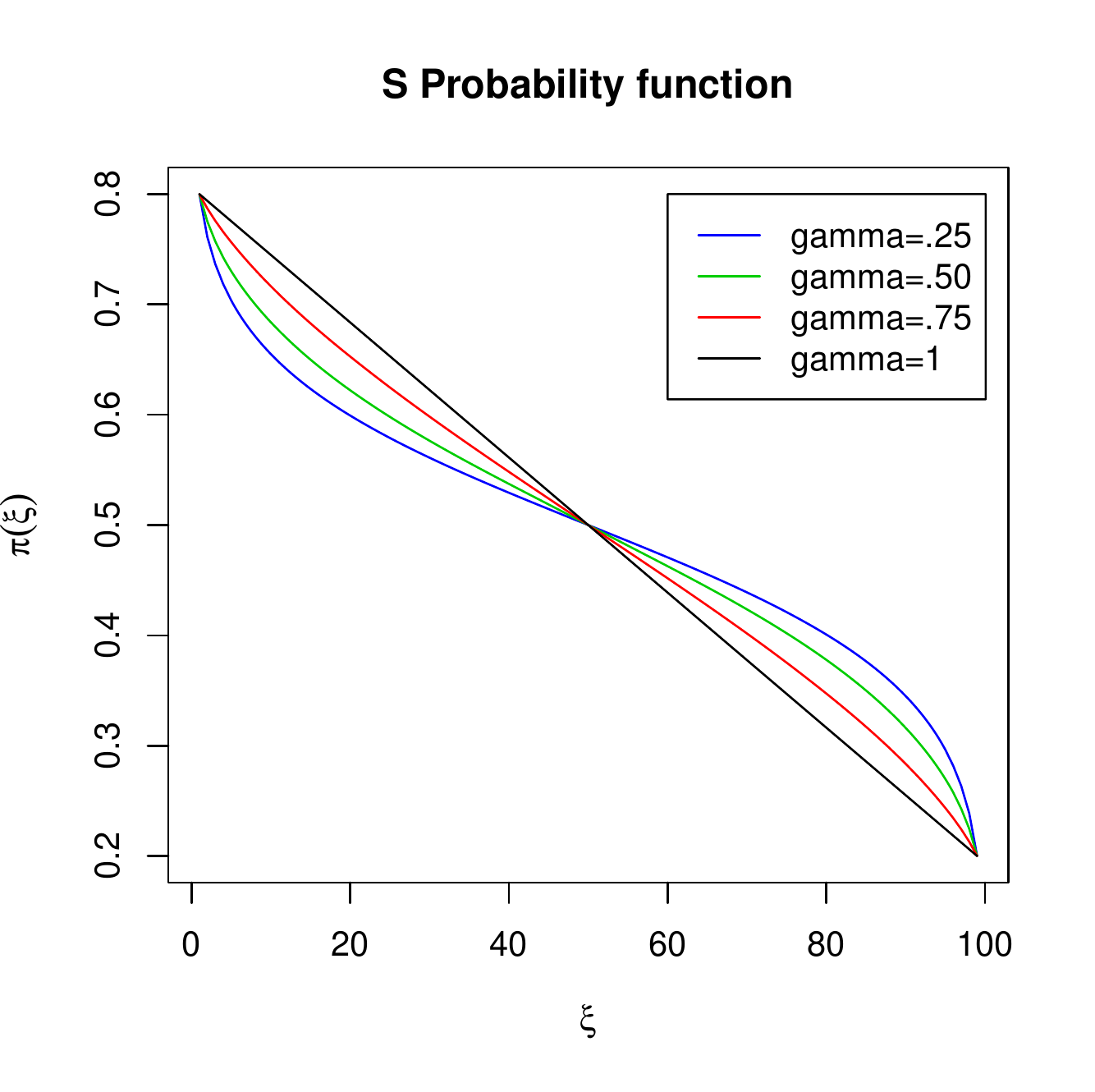} 
\end{tabular}
\end{figure}

The ``race level'' $\xi$ is the intervention in our experiment. The motivation behind designing the experiment comes from the need for investigating how the probability $\pi(\xi)$ of a receiver responding to a query from a sender with race level $\xi$ behaves as a function of $\xi$. Assuming that $\pi(\xi)$ is a monotonic function of $\xi$, interest lies in assessing whether there is an evidence of $\pi(\xi)$ increasing or decreasing monotonically with $\xi$, and if so, whether the function is linear, or non-linear. A monotonically decreasing function might seem unlikely to some, since it would imply that response rates decline as the race level changes from more likely black to more likely white, but for the subset of $\mathcal{R}$ that consists of black lawyers such a behavior of the function $\pi(\xi)$ is easily imagined.  If the function $\pi(\xi)$ is non-linear, then the interest lies in discriminating primarily between the two types of functions shown in Figures \ref{fig:two_functions1} (increasing in $\xi$) and \ref{fig:two_functions2} (decreasing in $\xi$). The left panel of Figure \ref{fig:two_functions1} shows a set of logistic functions gradually converging to a linear function as the underlying parameter $\gamma$ reduces to one. The right panel shows a set of inverted-S functions also approaching a linear function as $\gamma$ increases to one. Figure \ref{fig:two_functions2} shows similar functions that decrease with $\xi$. The functional form parametrized by $\gamma$ will be introduced in equations (\ref{eq:alphabeta})--(\ref{eq:b}).

Each response function $\pi(\xi)$ tells a different story about how racially disparate outcomes found in existing
correspondence studies relate to response rates for those with less strongly identified names. In the inverted S model
(right panel), individuals whose names are somewhat identified with a given race receive fairly similar outcomes to one
another, regardless of whether their name is more likely white or more likely black. In the logistic model (left panel), it
makes a bigger difference whether one is somewhat likely to be white or somewhat likely to be black. These differences
between the two probability functions increase as $\gamma$ moves further from one.

These differences also have real policy implications. If the logistic model reasonably represents the underlying behavior, the racial inequality identified by existing correspondence studies is larger and more pervasive than one might have assumed, for example via linearly interpolating between the two extremes. The difference shown in existing studies between black and white response rates would come closer to the difference experienced between the typical black and typical white individual. If the inverted S model reasonably represents the underlying behavior, the inequalities that correspondence studies have found is smaller and more concentrated than linear interpolation would suggest. The difference in response rates for the typical Caucasian and the typical African-American would be closer to zero. Given the latter finding, future research would ideally focus more narrowly on understanding the stereotypes that lawyers and other service providers have about individuals with distinctively black names. Policy interventions aimed at combating these stereotypes might be enough. By contrast, if even modest hints that a person is black trigger massively disparate treatment, then interventions challenging stereotypes are unlikely to suffice.

To further formulate this inference objective from this experiment conducted with a set $\mathcal{R}$ of $N$ receivers (units) with similar characteristics, with the $i$th receiver receiving treatment level $\xi_i$ (that is, a query from a sender whose name has a race-level $\xi_i$), we assume that the $i$th unit generates a binary response $Y_i$, which is one if he/she responds to the email and zero if not. Further, we assume that $Y_1, \ldots Y_N$ are independent Bernoulli random variables with $P[Y_i =1] = \pi(\xi_i)$ for $i = 1, \ldots N$. The function $\pi(\xi)$ satisfies boundary conditions
\begin{equation}
\pi(0^+) = \alpha,  \quad \pi(1^-) = \beta, \label{eq:alphabeta}
\end{equation}
where $\pi(0^+) = \lim_{\xi \rightarrow 0^+} \pi(\xi)$ and $\pi(1^-) = \lim_{\xi \rightarrow 1^-} \pi(\xi)$ and $\alpha$ and $\beta$ are in $(0,1)$. To discriminate between the monotonically non-decreasing logistic and inverted-S functional forms of $\pi(\xi)$, we assume the following functional form of $\pi(\xi)$:
\begin{equation}
\pi(\xi) = \left \{ \begin{array}{cc}
\left\{a + g(\gamma, \xi) \right\}/(a+b), & \mbox{if} \ \alpha < \beta, \\ 
\left\{b - g(\gamma, \xi) \right\}/(a+b), & \mbox{if} \ \alpha > \beta, \\
\alpha & \mbox{if} \ \alpha = \beta, \label{eq:pi_function}
\end{array}
\right.
\end{equation}
where for $\gamma >0$, $a > 0$ and $b > 1$,
\begin{eqnarray}
g(\gamma, \xi) &=& \xi^{\gamma}  / \left\{ \xi^\gamma + (1-\xi)^{\gamma} \right\},  \label{eq:g_function} \\
a &=&  \left\{ \begin{array}{cc} \alpha/(\beta-\alpha), & \alpha < \beta \\
(1-\alpha)/ \left(\alpha - \beta \right) & \alpha > \beta
\end{array} \right. \label{eq:a} \\
b &=& \left \{  \begin{array}{cc} \left(1- \alpha \right) \left(\beta - \alpha \right), & \alpha < \beta \\ 
\alpha / \left(\alpha - \beta \right) & \alpha > \beta
\end{array} \right.\label{eq:b}
\end{eqnarray}
Note that $g(\xi, \gamma)$ satisfies $g(\gamma, 0^+) = 0$ and $g(\gamma, 1^-) = 1$. Also, (\ref{eq:a}) and (\ref{eq:b}) ensure that $\pi(\cdot)$ in (\ref{eq:pi_function}) satisfies (\ref{eq:alphabeta}).  

The function $g(\gamma, \xi)$, its variants and more generalized forms have found applications in psychology literature for weighting probability functions \citep[see, for example][]{Goldstein1987, Gonzalez1997}. The function is inverted-S for $\gamma < 1$, linear for $\gamma = 1$ and logistic for $\gamma >1$. The shapes in Figure \ref{fig:two_functions1} were generated from (\ref{eq:pi_function})--(\ref{eq:b}) by setting $\alpha=0.2$, $\beta=0.8$ and by substituting $\gamma = 1, 1.25, 1.50, 1.75$ (left panel) and $\gamma = 0.25, .50, .75, 1$ (right panel). The shapes in Figure \ref{fig:two_functions2} were generated by reversing the values of $\alpha$ and $\beta$ to 0.8 and 0.2 respectively, keeping all other settings the same as in Figure \ref{fig:two_functions1}.

The problem of drawing inference on the function $\pi(\cdot)$ can be considered equivalent to the problem of inferring the parameter vector $(\alpha, \beta, \gamma)^{\T}$ or ${\bm \theta} = (a, b, \gamma)^{\T}$ after re-parameterization, where $\T$ denotes transposition. In Section \ref{Sec:MLE_Fisher}, we develop such an inference procedure assuming that the experimental data is available.

\section{Statistical inference of model parameters} \label{Sec:MLE_Fisher}

The results derived in this Section are based on the asymptotic theory of maximum likelihood estimation, that can be found in most textbooks on statistical inference \citep[see, for example][Chapter~6]{Boos2013}.
Let $\xi_i$ denote the level of the treatment assigned to unit $i \in \{1, \ldots, N\}$, and let $y_i$ denote the binary response obtained from unit $i$. Recall from Section \ref{Sec:stat_formulation}, that the $y_i$'s are assumed to be realizations of independent Bernoulli random variables $Y_i$'s with $P[Y_i = 1] = \pi(\xi_i)$ where $\pi(\xi_i)$ is given by (\ref{eq:pi_function})--(\ref{eq:b}). For ease of exposition, we assume that $\pi(\cdot)$ is decreasing, such that $\alpha > \beta$. As will be seen later in Section \ref{s:analysis}, the experimental data by and large indicate this pattern. The case of increasing $\pi(\cdot)$ can be handled exactly in the same way. The likelihood function of the parameter vector ${\bm \theta} = (a, b, \gamma)$ given the data $({\bm \xi},{\bm y})$, where ${\bm \xi} = (\xi_1, \ldots, \xi_N)^\T$ and ${\bm y} = (y_1, \ldots, y_N)^\T$ is
\begin{eqnarray*}
L({\bm \theta} | {\bm \xi}, {\bm y}) &\propto& \prod_{i=1}^N \left\{ \pi(\xi_i) \right\}^{y_i} \left\{ 1 - \pi(\xi_i) \right\}^{1 - y_i} \\ &\propto& \prod_{i=1}^N \left[ \left\{b - g(\gamma, \xi) \right\}/(a+b) \right]^{y_i} \left[ \left\{a + g(\gamma, \xi) \right\}/(a+b) \right]^{1 - y_i},  
\end{eqnarray*}
where $a$, $b$ and $g(\gamma, \xi)$ are defined in (\ref{eq:a}), (\ref{eq:b}) and (\ref{eq:g_function}) respectively. The second line follows by substituting the expression for $\pi(\xi_i)$ from (\ref{eq:pi_function}) corresponding to the case $\alpha > \beta$. Taking the logarithm, after some simplifications, the log-likelihood function is obtained as
\begin{equation}
\ell({\theta} | {\bm \xi}, {\bm y}) = \sum_{i=1}^N \left[ y_i \log \left( b - g(\gamma, \xi_i) \right) + (1-y_i) \log \left( a + g(\gamma, \xi_i) \right) \right] - N \log(a+b). \label{eq:loglike}
\end{equation}
The MLE $\widehat{\bm \theta}_N$ of the parameter vector ${\theta}$ based on a sample size of $N$ can be obtained by maximizing (\ref{eq:loglike}). The gradient of the log-likelihood, $\nabla \ell({\bm \theta})$ is given by the vector
\begin{eqnarray}
\left( \begin{array}{c}
\sum_{i=1}^N (1 - y_i) / \left\{ a + g(\gamma, \xi) \right\} - N/(a+b) \\
\sum_{i=1}^N y_i / \left\{ b - g(\gamma, \xi) \right\} - N/(a+b) \\
\sum_{i=1}^N  g^{\prime}(\gamma, \xi_i) \left \{ (1 - y_i)/\left(a + g(\gamma, \xi_i)\right) - y_i/\left(b - g(\gamma, \xi_i) \right) \right\}
\end{array}
\right)  \label{eq:nabla_theta}
\end{eqnarray}
where 
\begin{eqnarray}
g^{\prime}(\gamma, \xi_i) = \frac{\partial g(\gamma, \xi_i)}{\partial \gamma} =\frac{\partial} {\partial \gamma} \left \{ \frac{\xi_i^{\gamma}}{\xi_i^{\gamma} + (1-\xi_i)^{\gamma}} \right \} =  \frac{ \{\xi_i (1 - \xi_i)\}^{\gamma} }{\{\xi_i^{\gamma} + (1-\xi_i)^{\gamma} \}^2 } \log \left( \frac{\xi_i}{1 - \xi_i} \right). \label{eq:g_prime}
\end{eqnarray}

Equating $\nabla \ell({\bm \theta})$ to zero yields the maximum likelihood estimator of ${\bm \theta}$. However, since we do not have a closed-form solution of the ML equations, we use the following iterative algorithm to obtain the MLE:

\noindent\rule{16cm}{1pt}
\begin{center}{ALGORITHM FOR OBTAINING MLE OF ${\bm \theta}$} \end{center}
\begin{itemize}
\item INITIALIZATION
\begin{itemize}
\item[Step 1:] Obtain preliminary estimators of $\alpha$ and $\beta$ as $\widehat{\alpha}^{(0)} = \overline{y}(\xi_{\min})$ and $\widehat{\beta}^{(0)} = \overline{y}(\xi_{\max})$, where $\overline{y}(\xi_{\min})$ and $\overline{y}(\xi_{\max})$ respectively denote the average outcomes (or observed proportions of responses from receivers) at the minimum ($\xi_{\min}$) and maximum ($\xi_{\max}$) levels of $\xi$ used in the experiment. The subsequent steps assume that $\hat{\alpha}^{(0)} > \hat{\beta}^{(0)}$ (if not, at each step the corresponding expressions and functions for the case $\hat{\alpha}^{(0)} < \hat{\beta}^{(0)}$ will be used.
\item[Step 2:] Obtain preliminary estimators of $a$ and $b$ by substituting $\hat{\alpha}^{(0)}$ and $\hat{\beta}^{(0)}$ for $\alpha$ and $\beta$ in (\ref{eq:a}) and (\ref{eq:b}). That is, obtain  $\widehat{a}^{(0)} = \left(1 - \widehat{\alpha}^{(0)} \right)/ \left( \widehat{\alpha}^{(0)} - \widehat{\beta}^{(0)}  \right)$ and $\widehat{b}^{(0)} = \widehat{\alpha}^{(0)} / \left( \widehat{\alpha}^{(0)} - \widehat{\beta}^{(0)}  \right)$.
\item[Step 3:] Obtain the preliminary estimator of $\gamma$ as:
$\widehat{\gamma}^{(0)} = \arg \underset{\gamma}{\max} \ \ell(\widehat{a}^{(0)},\widehat{b}^{(0)},\gamma | {\bm \xi}, {\bm y}),$ where $\ell(a, b, \gamma)$ is given by (\ref{eq:loglike}).
\end{itemize}
\item ITERATION: Assuming we have estimators $\widehat{\bm \theta}^{(t)}$, update estimators of $a$, $b$ and $\gamma$ recursively using the following steps:
\begin{itemize}
\item[Step 6:] Obtain $\widehat{a}^{(t+1)}$ by solving
$ \sum_{i=1}^N (1 - y_i) / \left\{ a + g(\gamma^{(t)}, \xi) \right\} - N/(a+\widehat{b}^{(t)}) = 0$ for $a$.
\item[Step 7:] Obtain $\widehat{b}^{(t+1)}$ by solving
$ \sum_{i=1}^N y_i / \left\{ b - g(\gamma^{(t)}, \xi) \right\} - N/(\widehat{a}^{(t)}+b) = 0$ for $a$.
\item[Step 8:] Update the estimate of $\gamma$ by solving the one-dimensional optimization problem:
$$ \widehat{\gamma}^{(t+1)} = \arg \underset{\gamma}{\max} \ \ell(\widehat{a}^{(t+1)},\widehat{b}^{(t+1)},\gamma | {\bm \xi}, {\bm y}),$$ where $\ell(a, b, \gamma)$ is given by (\ref{eq:loglike}).
\end{itemize}
\end{itemize}

\bigskip

\textbf{Stopping rule}: Terminate the algorithm if 
$$ \max \{ |\widehat{a}^{(t+1)} - \widehat{a}^{(t)}|, |\widehat{b}^{(t+1)} - \widehat{b}^{(t)}|, |\widehat{\gamma}^{(t+1)} - \widehat{\gamma}^{(t)}| \} < \epsilon,$$ where $\epsilon > 0$ is a pre-defined threshold.

\noindent\rule{16cm}{1pt}


The following proposition (proven in the Appendix) gives the asymptotic distribution of $\widehat{\theta}_N$. Note that in the proposition, we use the notation $\overset{.}{\sim}$ to denote ``approximately distributed as''.

\begin{proposition} \label{prop:asym_theta}
For large $N$, 
\begin{equation}
\widehat{\theta}_N  \overset{.}{\sim} \mathcal{N} \left({\bm \theta}, \left\{{\mathbf I}_N({\bm \theta})\right\}^{-1} \right), \label{eq:MLE_dist}
\end{equation}
where ${\mathbf I}_N ({\bm \theta}) = \sum_{i=1}^N {\mathbf I}_i ({\bm \theta})$ is the expected Fisher information matrix obtained from $N$ data points $({\bm \xi}, {\bm y})$, with
\begin{equation}
\mathbf{I}_i({\bm \theta}) = \left( 
\begin{array}{ccc}
\frac{b - g(\gamma, \xi_i)}{(a+b)^2 \left\{ a + g(\gamma, \xi_i) \right\}} & -\frac{1}{(a+b)^2} & \frac{g^{\prime}(\gamma, \xi_i)}{(a+b) \left\{ a + g(\gamma, \xi_i) \right\}} \\
-\frac{1}{(a+b)^2} & \frac{a + g(\gamma, \xi_i)}{(a+b)^2 \left\{ b - g(\gamma, \xi_i) \right\}} & -\frac{g^{\prime}(\gamma, \xi_i)}{(a+b) \left\{ b - g(\gamma, \xi_i) \right\}} \\
\frac{g^{\prime}(\gamma, \xi)}{(a+b) \left\{ a + g(\gamma, \xi_i) \right\}} & -\frac{g^{\prime}(\gamma, \xi_i)}{(a+b) \left\{ b - g(\gamma, \xi_i) \right\}} &  \frac{\left( g^{\prime}(\gamma, \xi_i) \right)^2}{\left\{ a + g(\gamma, \xi_i) \right\} \left\{ b - g(\gamma, \xi_i) \right\}}, 
\end{array}
\right)
\end{equation}  \label{eq:MLE_Avar}
where $g^{\prime}(\gamma, \xi_i)$ is given by (\ref{eq:g_prime}).
\end{proposition}

Note that Proposition \ref{prop:asym_theta} cannot be directly used to construct asymptotic confidence intervals or conduct tests of hypothesis for ${\bm \theta}$ as the asymptotic covariance matrix term depends on ${\bm \theta}$. However, we can use a plug-in estimator of the approximate covariance matrix by substituting the MLE $\widehat{\bm \theta}_N$ in place of ${\bm \theta}$ in the expected Fisher information $\mathbf{I}_N({\bm \theta})$.

Whereas Proposition \ref{prop:asym_theta} provides us with the asymptotic covariance matrix of the MLE of ${\bm \theta} = (a, b, \gamma)^{\T}$, it is important to obtain the covariance matrix of the original parameters $\alpha$ and $\beta$ to determine whether the probability functions are increasing or decreasing. The following corollary provides this result.

\begin{corollary} \label{cor:alphabeta_var}
For large $N$, the approximate covariance matrix of the ML estimators of $\alpha$ and $\beta$ is $\mathbf{H}^{\T} \Sigma_{ab} \mathbf{H}$, where $\Sigma_{ab}$ is the $2 \times 2$ principal submatrix obtained by removing the third row and third column of $\left\{\mathbf{I}_N({\bm \theta})\right\}^{-1}$ defined in Proposition \ref{prop:asym_theta}  and
\begin{eqnarray*}
\mathbf{H} = \left( \begin{array}{cc} -b/(a+b)^2 &  -(b-1)/(a+b)^2 \\    
a/(a+b)^2 & (a+1)/(a+b)^2
\end{array} \right)
\end{eqnarray*}
\end{corollary}

\bigskip

The last result of this section (Corollary \ref{cor:pi_var}) provides an expression for the asymptotic variance of the predicted probability function $\pi(\xi)$ at an arbitrary race level. Such a prediction can be made by substituting the MLEs of $a$, $b$ and $\gamma$ in the appropriate expression in (\ref{eq:pi_function}).
\begin{corollary} \label{cor:pi_var}
For large $N$, the approximate variance of the predicted probability $\widehat{\pi}(\xi_{\new})$ at a new race level $\xi_{\new}$ is ${\bm \lambda}_{\new} \ \{ \mathbf{I}_N({\bm \theta}) \}^{-1}  \ {\bm \lambda_{\new}}$, where 
\[ {\bm \lambda}_{\new}^{\T} = \left( \begin{array}{c} 
\{ b - g(\gamma, \xi_{\new}) \}/(a+b)^2 \\
\{ a + g(\gamma, \xi_{\new}) \}/(a+b)^2 \\ 
 g^{\prime}(\gamma, \xi_{\new})/(a+b)
\end{array}
\right)
\]
\end{corollary}

\section{Designing the experiment} \label{Sec:Design}

As discussed in Section \ref{sec:intro}, there are four key questions to answer while designing the experiment: (a) how many levels ($k$) of the treatment ``race level'' should we select? (b) How to identify names with the intended ``race levels''? (c) What should be the minimum number of experimental units $N$ to guarantee reasonable precision of the inference?  (d) How should one allocate $N$ available experimental units (lawyers) to the $k$ treatment levels? In the following three subsections, we describe how questions (a), (b) and (d) were answered and the answers implemented in the current experiment. The answer to (c), i.e., selection of $N$ was essentially guided by resource constraints explained in Section \ref{ss:blocks}. However, we present some ideas to determine the adequacy of the sample size in Appendix \ref{sec:chooseN}, which also suggests that the available resources were probably inadequate to draw meaningful inference from our experimental data.

\subsection{An optimal design approach to determine the number of treatments} \label{ss:optimal_design}

A common way to address such problems is to use an optimal design formulation \citep{Atkinson2007}, in which one can consider an information-based design criterion, such as the determinant or trace of the Fisher information matrix. Optimization of the determinant leads to a D-optimal design, which is a popular criterion for designing optimal experiments. However, the model under consideration is non-linear, making the design criterion dependent on the parameter. Therefore, the optimal design solution is dependent on the true value of the parameter ${\bm \theta}$, typically known as the local optimal design \citep{Chernoff1953}. In such cases to find the optimal design, one has to substitute a ``guess'' of the true parameter value, which defeats the basic purpose of our study. A popular alternative to this approach is to use a sequential strategy, which may be either frequentist \citep{Chaudhuri1993} or Bayesian \citep[see, for example][]{Chaloner1995, Zhu2014}. However, in our case, designing a sequential experiment was not possible due to resource, time and logistics-related issues. Further, it has to be kept in mind that in reality the data may reveal a functional form of $\pi(\xi)$ that is different from the one described by (\ref{eq:pi_function})--(\ref{eq:b}). 
Thus, instead of taking a ``hard optimization'' route, we decided to adopt a ``softer'' space-filling design strategy while retaining the spirit of a D-optimal design by conducting the experiment with $k$ equally spaced levels of $\xi$, each of which would be assigned to equal number $r=N/k$ of experimental units. The problem would then be to determine $k$ such that the D-optimality criterion defined above is reasonably large across a wide range of parameter settings. 

Armed with the results in Section \ref{Sec:MLE_Fisher} we conducted extensive simulations with different parameter settings to identify $k$ that maximizes the D-optimality criterion. In the simulations, the number of experimental units $N$ was kept fixed at 1200. Note that changing $N$ does not alter the optimality patterns, as the problem is essentially to find the number of levels over which to distribute $N$ equally. However, we need to fix $N$ to compute the D-optimality criterion $\text{det} \left( {\mathbf I}_N ({\bm \theta}) = \sum_{i=1}^N {\mathbf I}_i ({\bm \theta}) \right)$, where ${\mathbf I}_i ({\bm \theta})$ is given by (\ref{eq:MLE_Avar}). The $k$ treatment levels $\xi_1, \ldots, \xi_k$ were obtained as 
\begin{equation}
\xi_j = .01 + .98 (j-1)/(k-1), \ j=1, \ldots, k. \label{eq:xi_j}
\end{equation}
For each setting of the parameter values $\alpha, \beta, \gamma$, the following nine levels of $k$ were chosen: 3, 4, 5, 6, 8, 10, 12, 15, 20. These levels of $k$ ensured that $r = N/k = 1200/k$ representing the number of units receiving each treatment was an integer.   

\begin{figure}[ht]
\centering
\caption{$D$-optimality criteria versus levels of $k$ when $\beta - \alpha \le 0.3$} \label{fig:D-optimality1}
\begin{tabular}{cc}
\includegraphics[scale=0.30]{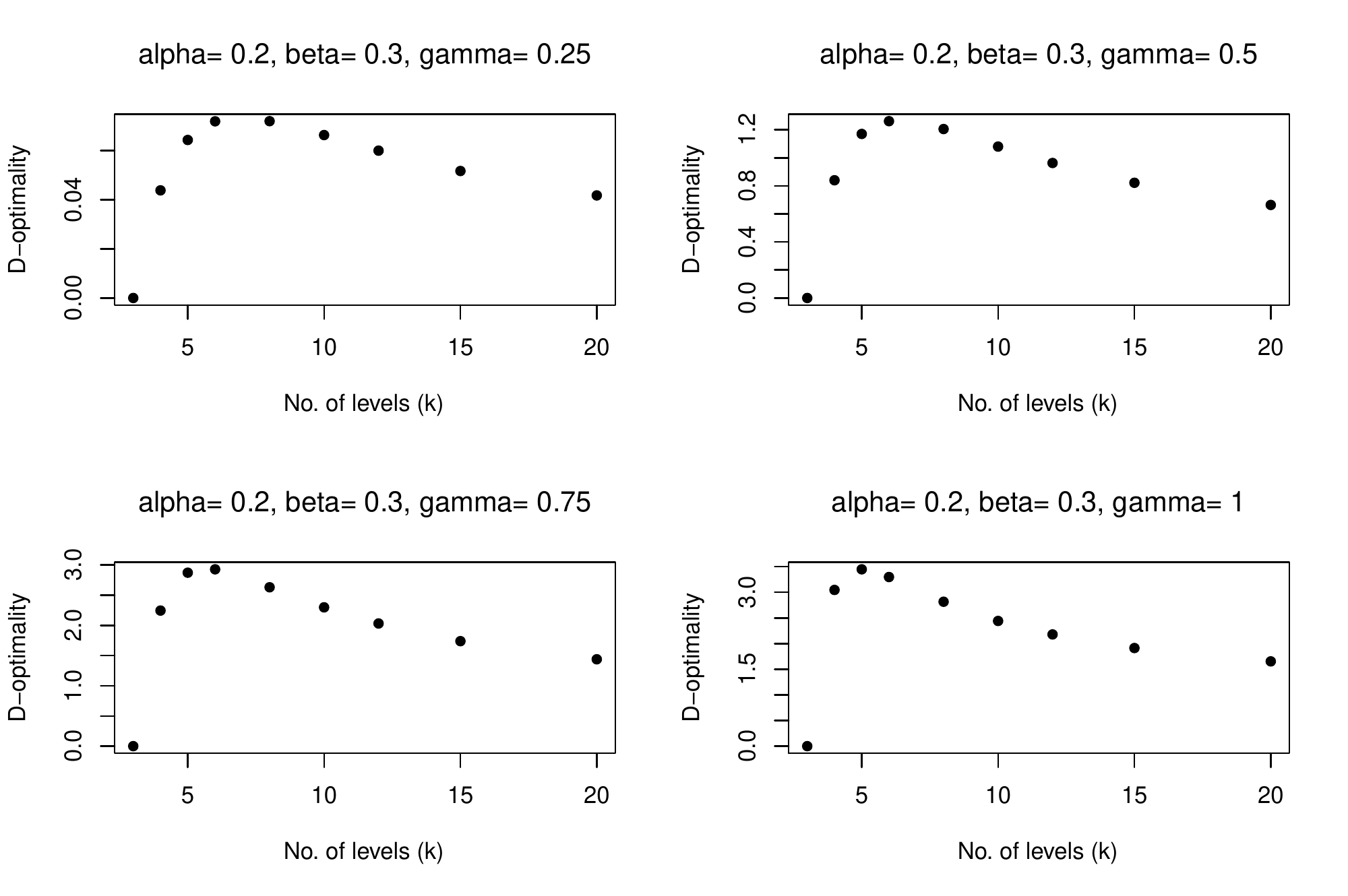} &
\includegraphics[scale=0.30]{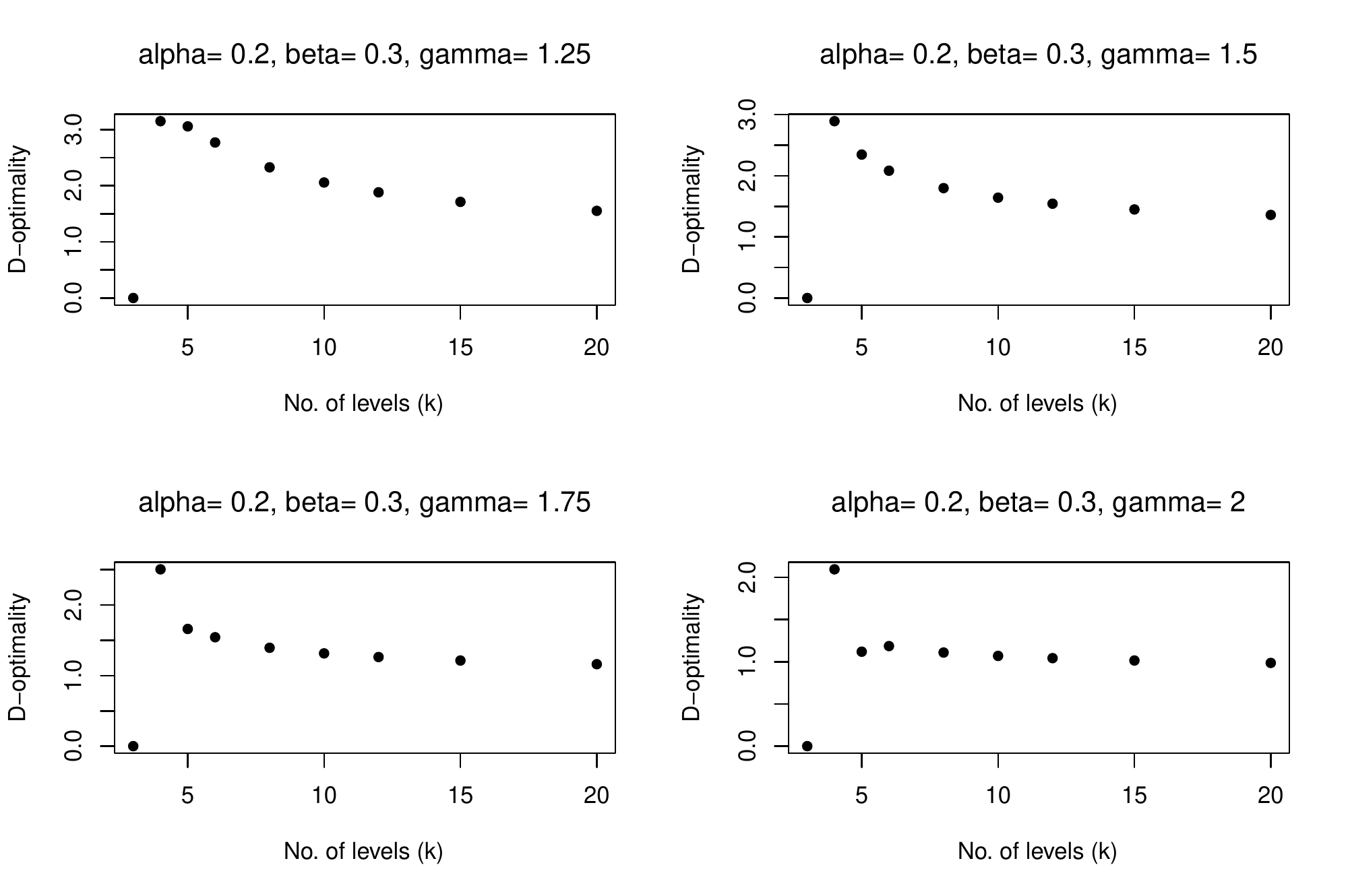} \\
\includegraphics[scale=0.30]{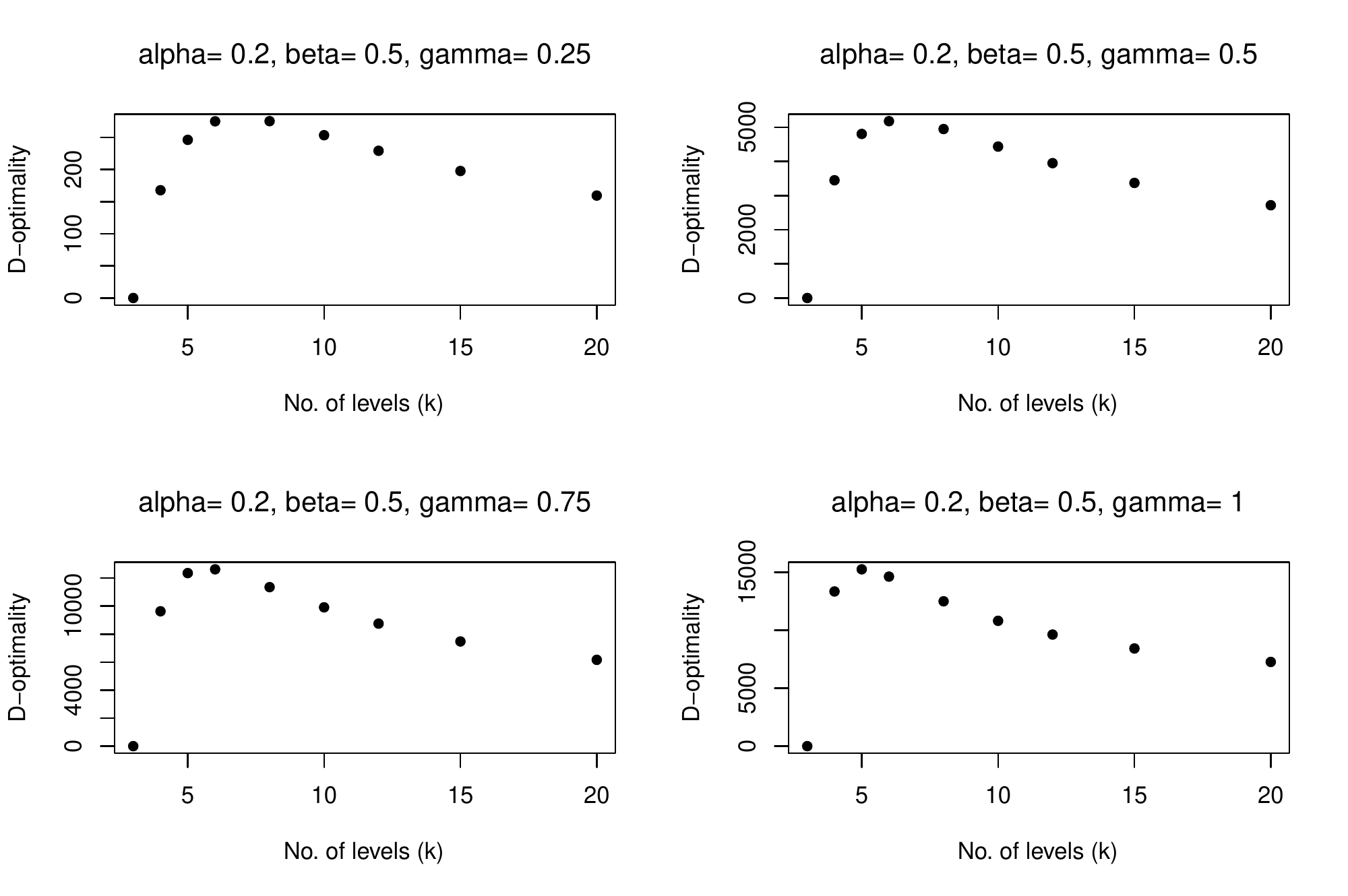} &
\includegraphics[scale=0.30]{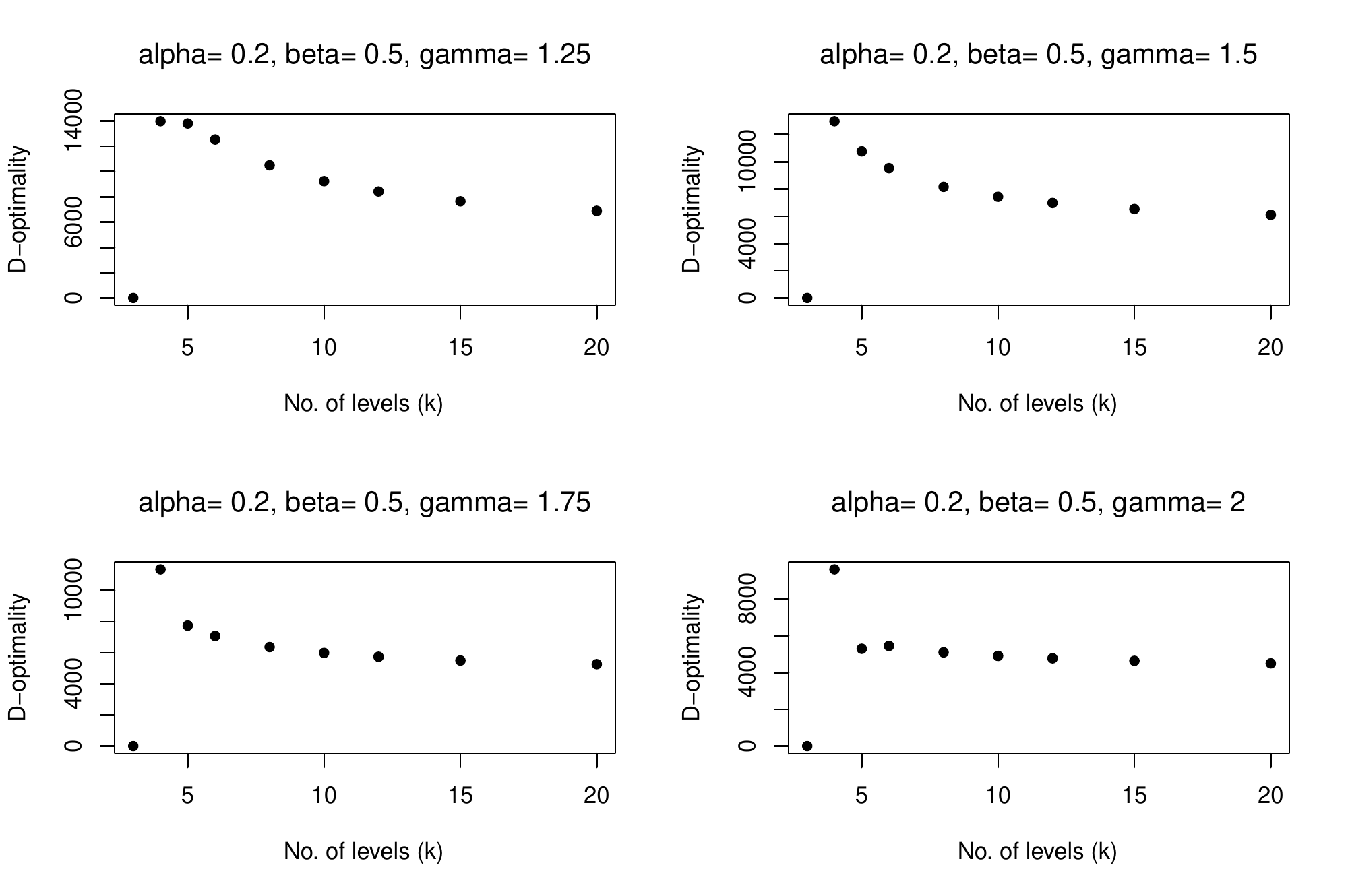} \\ 
\end{tabular}
\end{figure}

\begin{figure}[ht]
\centering
\caption{$D$-optimality criteria versus levels of $k$ when $\beta - \alpha > 0.3$} \label{fig:D-optimality2}
\begin{tabular}{cc}
\includegraphics[scale=0.30]{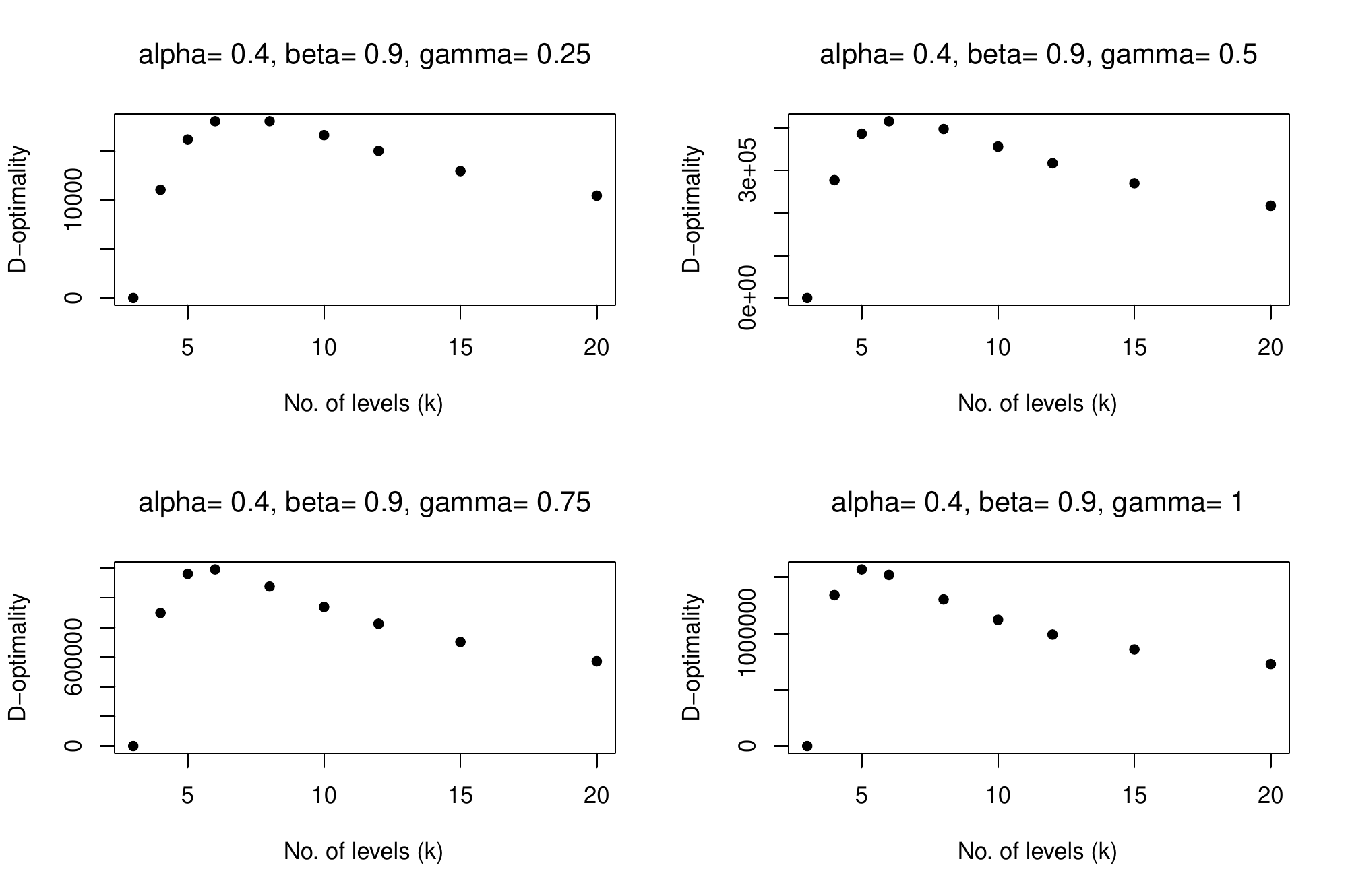} &
\includegraphics[scale=0.30]{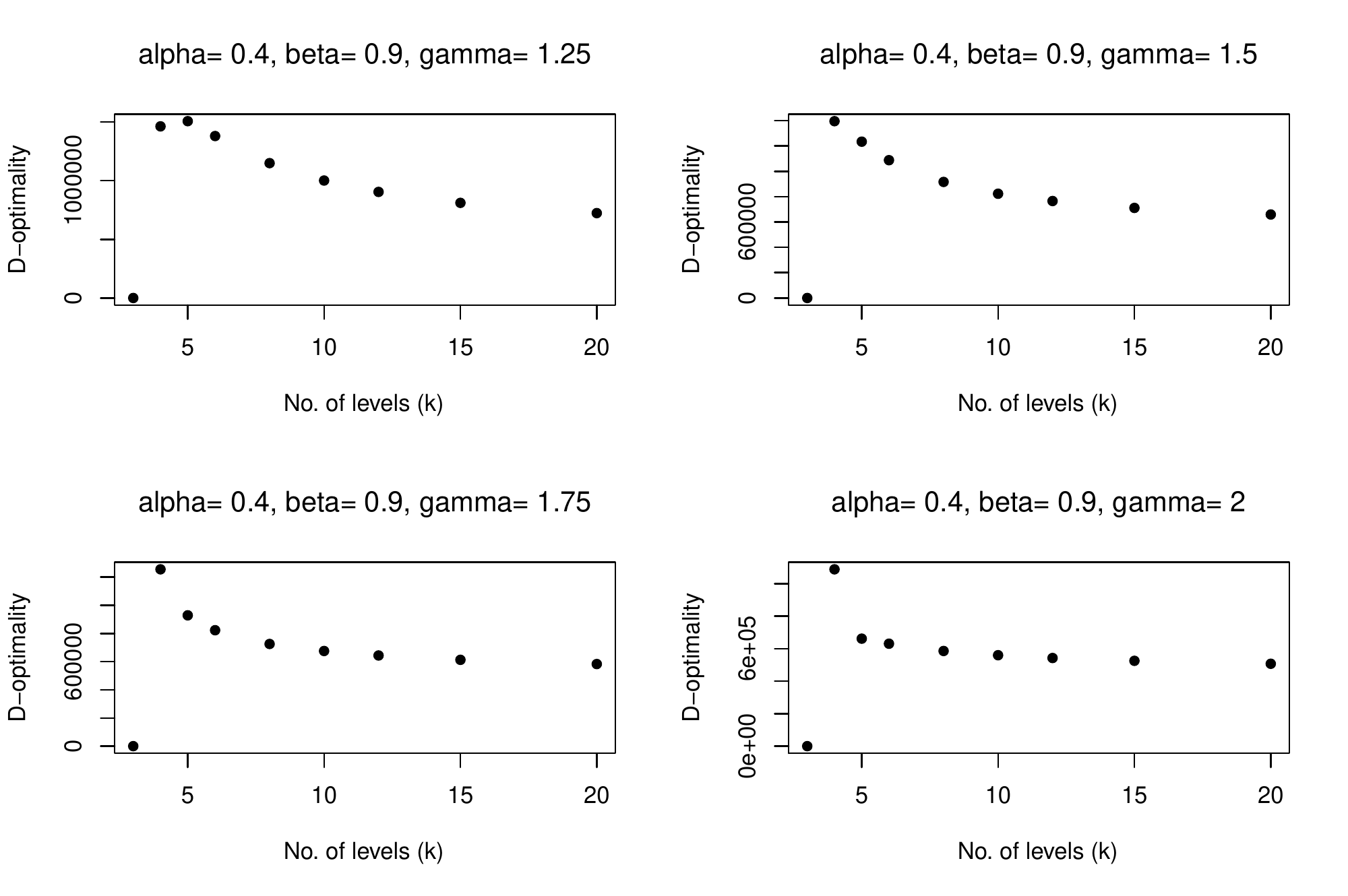} \\
\includegraphics[scale=0.30]{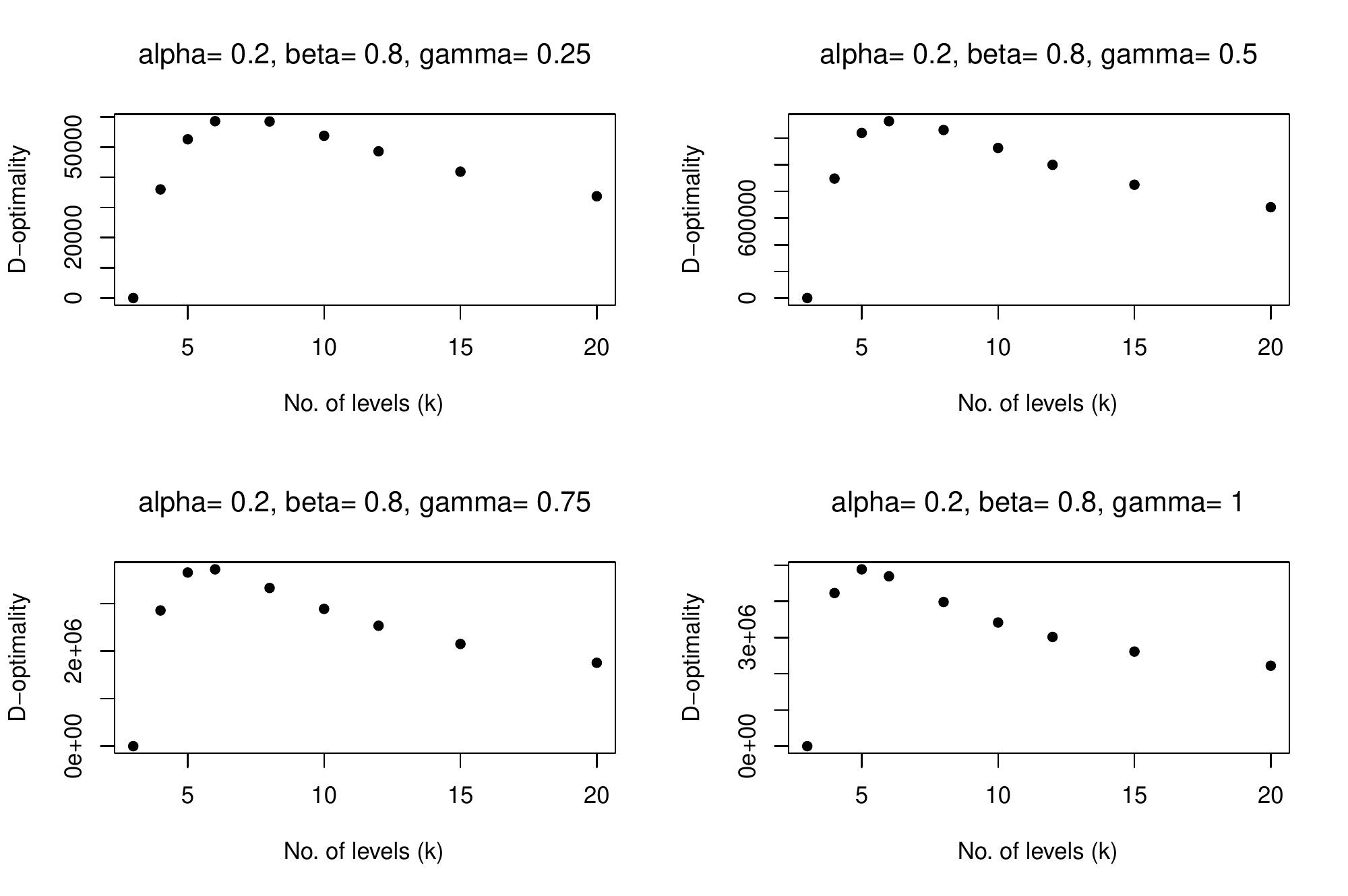} &
\includegraphics[scale=0.30]{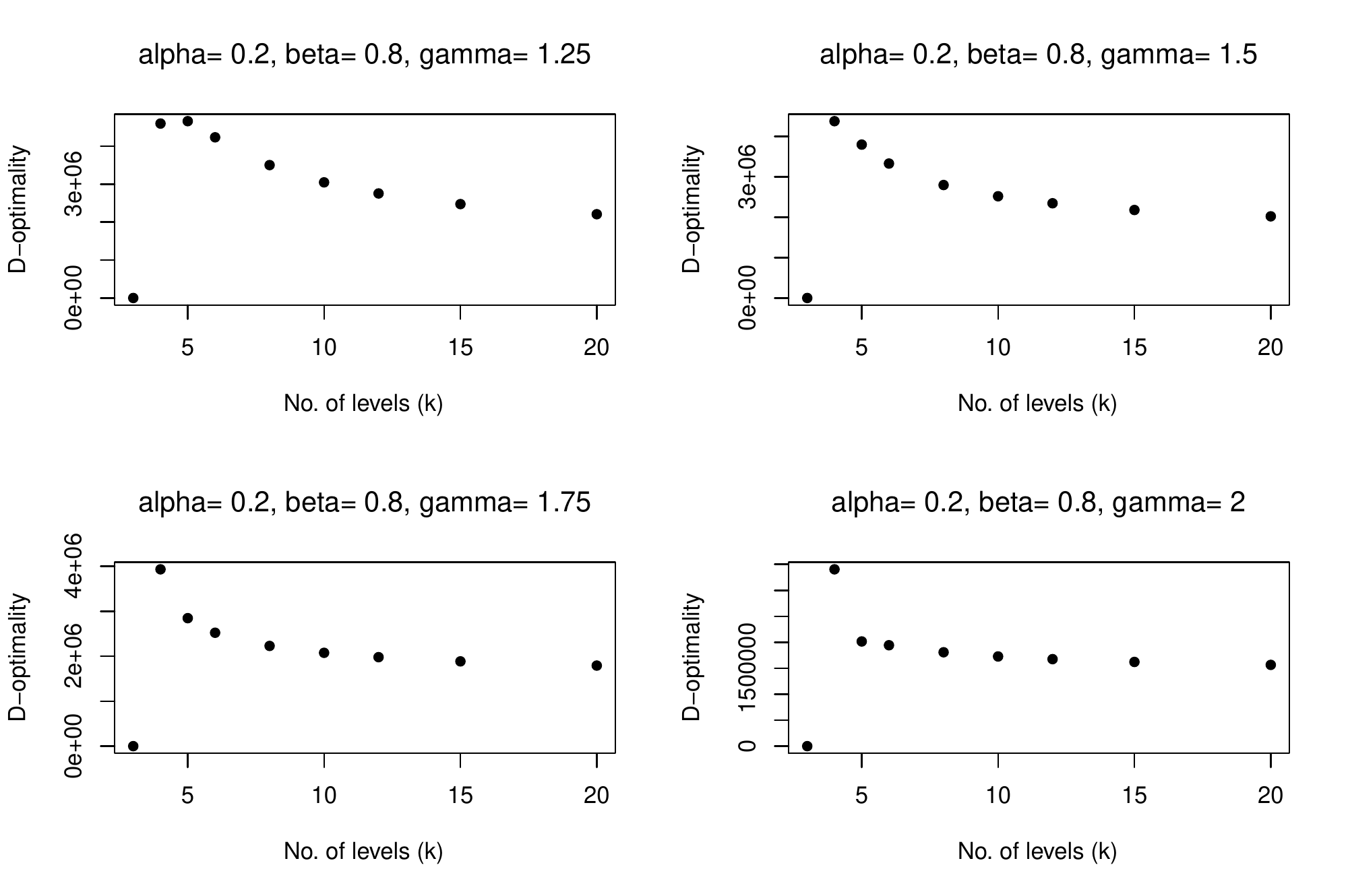} \\ 
\end{tabular}
\end{figure}

We chose the following combinations of parameters $\alpha$ and $\beta$: (0.2,0.3), (0.2,0.5), (0.4, 0.9), (0.2, 0.8) to cover a wide-range of possibilities where $\alpha$ and $\beta$ were close or distant. For each such combination, we considered true values of $\gamma$ ranging from 0.25 to 2 in increments of 0.25. For each fixed value of $\alpha$, $\beta$, $\gamma$, $k$ and $r = 1200/k$ we generated the vector ${\bm \xi}$ of length 1200 by replicating each level $\xi_1, \ldots, \xi_k$ defined in (\ref{eq:xi_j}) $r$ times. Finally, the D-optimality criterion was computed for each parameter combination and plotted across different levels of $k$. These plots are shown in Figure \ref{fig:D-optimality1} where $\beta - \alpha \le 0.3$ and Figure \ref{fig:D-optimality2} where $\beta - \alpha > 0.3$ in the Appendix. 

Both figures showed a similar pattern. For all combinations of $\alpha$ and $\beta$, six levels appeared to be the optimal choice when $\gamma < 1$, whereas four/five levels appeared to be the best choice when $\gamma$ exceeded one.  Overall, six levels appeared to be a reasonable choice that was robust across all choices of $\alpha$, $\beta$ and $\gamma$. Hence, it was decided to use six levels for the experiment and assign approximately the same number of units (lawyers) to each level.

\subsection{Choosing names with desired levels} \label{ss:name_loading}

Having made the decision to use six names that were racially identified to varying degrees, it was necessary to obtain names that had the intended race levels or odds of being white/black. To the best of our knowledge, no prior study has looked at names with intermediate racial signal levels.
  
In order to conduct this search, a large administrative dataset containing racially-identified name information was considered. In particular, the Florida voter file containing the first and last names of $M_w = 8.2$ million self-described whites and $M_b = 1.7$ million self-described African-Americans was used. Suppose at least one of these $M = M_w + M_b$ voters had the name $XY$ where $X$ denoted the first name and $Y$ the last name. The problem is now to calculate the probability that a voter chosen randomly from this population with name $XY$ is white (or black). To do this, we used a ``naïve Bayesian'' approach for setting the odds of a person’s race from their name similar to one used by \cite{Fryer2004}. 

Let $F_{Xw}$ and $F_{Xb}$ denote the numbers of white and black voters in the list with first name $X$. Similarly, let $L_{Yw}$ and $L_{Yb}$ denote the numbers of white and black voters with surname $Y$. Then, assuming that first names and last names are assigned independently, the probability that a white person chosen from the list at random will have name $XY$ is $(F_{Xw}/M_w)(L_{Xw}/M_w)$ and that a black person will have name will have name $XY$ is $(F_{Xb}/M_b)(L_{Xb}/M_b)$. Applying Bayes' theorem, the probability that a person chosen randomly from this list with name $XY$ is white is given by
\begin{equation}
\frac{(F_{Xw}/M_w)(L_{Yw}/M_w)(M_w/M)}{(F_{Xw}/M_w)(L_{Yw}/M_w)(M_w/M) + (F_{Xb}/M_b)(L_{Yb}/M_b)(M_b/M)}. \label{eq:white_odds}
\end{equation}

To illustrate the method, consider a name such as Terence Austin. According to the Florida voter file, 998 white Floridians have the given name Terence and 3,416 have the surname Austin. By contrast, 454 black Floridians have the given name Terence and 1554 have the name Austin. Substituting $F_{Xw} = 998$, $L_{Yw} = 3416$, $F_{Xb} = 454$, $L_{Yb} = 1554$, $M_w = 8227929$, $M_b = 1733249$ in (\ref{eq:white_odds}), the odds that Terence Austin would be white turns out to be 0.5044392.

The method described above provided a method of generating names that plausibly should signal a certain probability that a name belongs to a white person as opposed to a black person. Yet it was not immediately obvious that this objective estimate would correspond to subjective beliefs. In order to assess such correspondence, two large sample surveys were conducted on Amazon Mechanical Turk (m-Turk), one for each gender. For a comparable approach, see \cite{Gaddis2017} and \cite{Gaddis2017a}. The key idea was to pick two names (one male and one female) such that (a) subjective and objective probabilities of associating the name with a white or a black person were close and (b) the subjective probabilities of the six names for each gender were more or less equally spaced in the interval $[0,1]$. 

\begin{figure}[ht]
\centering
\caption{Female (left) and male (right) names identified for the experiment} \label{fig:twelve_names}
\begin{tabular}{cc}
\includegraphics[scale=0.45]{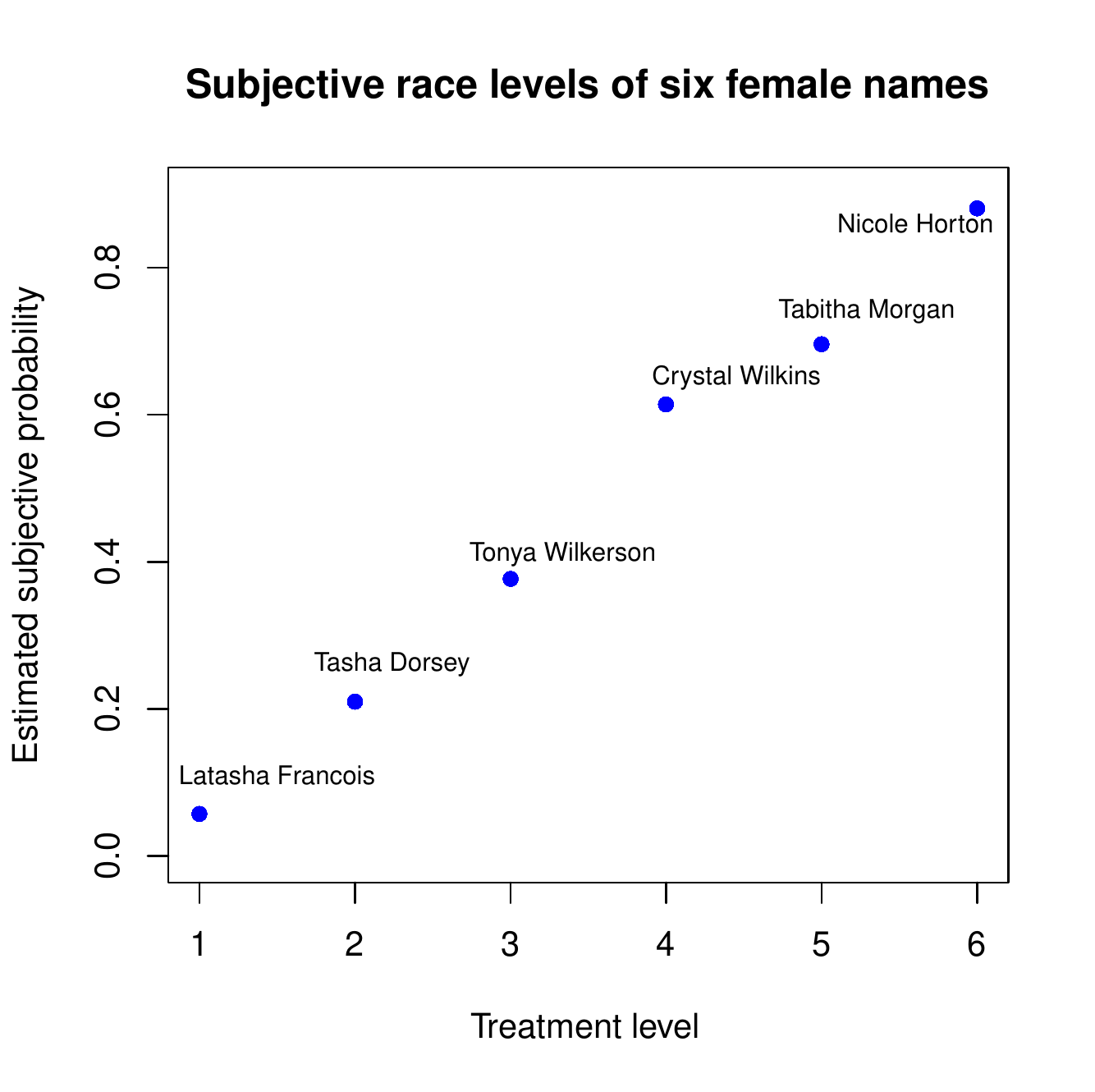} &
\includegraphics[scale=0.45]{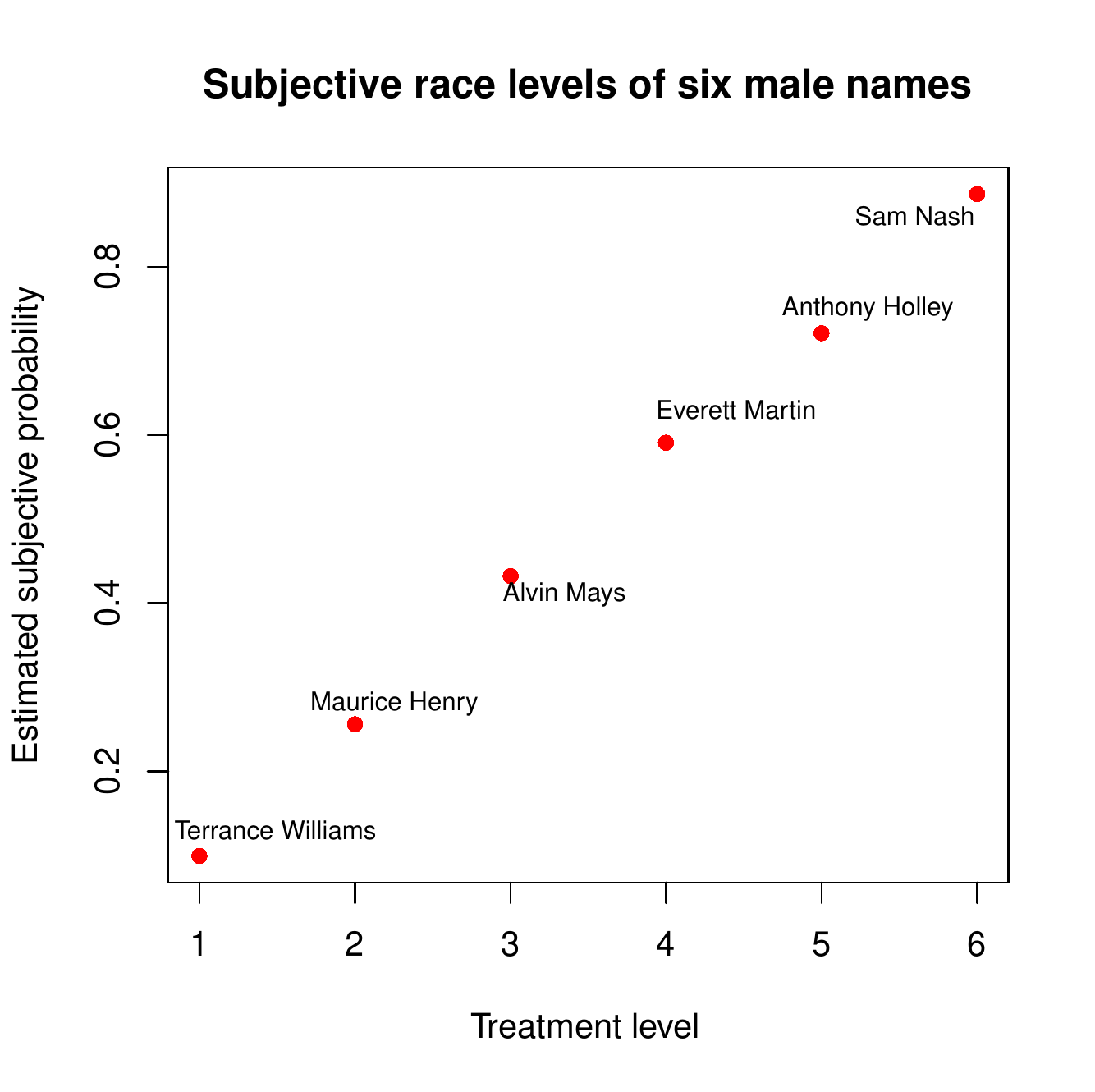} \\
\end{tabular}
\end{figure}

A set of names spanning more or less the entire spectrum of subjective race levels were shortlisted for this study. For each name, m-Turkers were asked to guess whether the individual with this name was more likely white or more likely black. The subjective odds of signaling white (or black) was estimated as the proportion of m-Turkers who identified it as white (or black). Eventually, six female names and six male names were identified for the study. These names along with their respective estimated subjective race levels are shown in Figure \ref{fig:twelve_names}.

\subsection{Experimental units, blocking and randomization} \label{ss:blocks}

The final and most costly aspect of the design of the experiment was to find adequate number of experimental units (lawyers) for inclusion in the study, and to identify their race, gender and practice area - considered as three important blocking factors that could possibly affect their potential reaction to the client's race. Only solo and small firm practitioners, understood as lawyers working in firms with fewer than six attorneys, were considered eligible for inclusion. A team of Harvard undergraduate research assistants was employed to skim through the Florida bar directory in random order and instructed to identify as many eligible lawyers as possible. The primary way they did so was through lawyer websites, although occasionally they might have identified a lawyer’s race through news coverage. In some cases, lawyers advertised themselves as practicing in several areas, for example criminal defense and personal injury. In these cases, a lawyer was considered eligible to fall into any block. With the available resources and time, a total of 899 lawyers were found for inclusion in the study. Table 1 shows the distribution of these 899 lawyers in each block.

\begin{table}[htbp]
\centering 
\caption{Lawyers in blocks}
\begin{tabular}{c|cc|cc}
Practice Area & \multicolumn{2}{c|}{Female} & \multicolumn{2}{c}{Male} \\
              & Black & White & Black & White                          \\ \hline
Criminal      & 48 & 49 & 74 & 97 \\ \hline
Divorce       & 73 & 72 & 49 & 72 \\ \hline
Personal Injury & 72 & 77 & 72 & 144 \\ \hline
\end{tabular}
\end{table}

Approximately half of the lawyers within each block were randomly assigned to female senders (that is, they would be receiving emails from potential female clients), while the other half were assigned to male senders. Lawyers assigned to female senders were further assigned to the six female names shown in the left panel of Fig \ref{fig:twelve_names} randomly in equal proportions. Similarly, Lawyers assigned to male senders were further assigned to the six male names shown in the right panel of Fig \ref{fig:twelve_names} randomly in equal proportions. Note that the six race levels were slightly different for female senders and male senders.

\section{Analysis} \label{s:analysis}

Based on the calculations shown in Appendix \ref{sec:chooseN}, the sample size of 899 would only be adequate to draw inference on the $\pi$-function with reasonable precision if (i) the magnitude of the treatment effect  $|\beta - \alpha|$ is large, (ii) $\gamma$ is far from 1, (iii) few (to none) of the block effects are significant, (iv) there are no unanticipated treatment interactions or environmental factors influencing these disparities. We considered it plausible that some of these conditions would fail, although which ones would and how badly was not obvious. Indeed, there seems to be good evidence that the legal market matters, a violation of (iv) not well anticipated by prior theory (see \cite{Libgober2020} for a longer discussion). However, we still present a demonstration of the recommended analysis consistent with the design of the experiment.

A preliminary analysis was done by calculating the percentage of responses received for each treatment level. Contrary to expectations, the observed proportion of responses at the lowest race level (very likely black) appeared to be slightly smaller than that at the highest race level (very likely white). Thus, the model was fit and parameter estimates obtained under the assumption that $\alpha < \beta$. 

Using the treatment levels and responses from the 899 lawyers, the maximum likelihood estimates of parameters $\alpha$, $\beta$ and $\gamma$ were obtained using the iterative algorithm described in Section \ref{Sec:MLE_Fisher}. The algorithm was terminated using the stated stopping rule with $\epsilon = 0.001$. The Fisher information matrix was computed by substituting these estimates in (\ref{eq:MLE_Avar}), and subsequently asymptotic variances of the parameters were obtained. The last two columns of Table \ref{tab:analysis_MLE} shows these estimates and their asymptotic standard errors. The standard error of $\widehat{\beta} - \widehat{\alpha}$ was obtained as $\sqrt{var(\widehat{\beta}) + var(\widehat{\alpha}) - 2 cov(\widehat{\beta},\widehat{\alpha})}$, where the variances and covariances could be obtained from the covariance matrix of $\widehat{\alpha}$ and $\widehat{\beta}$ from Corollary \ref{cor:alphabeta_var}. Using this standard error, a 95\% confidence interval for the treatment effect $\beta - \alpha$ was obtained as $[-0.4069, 0.7206]$. Because this interval included zero, it was not possible to reject the null hypothesis of zero treatment effect, making the question of identifying the shape of the function $\pi(\xi)$ (and consequently, estimation of $\gamma$) irrelevant. 

Because the race of the lawyer was thought to be the blocking factor with the largest potential impact on the response curves, we also analyzed the data separately for these two blocks, and the resuts are shown in Table \ref{tab:analysis_MLE}, again recognizing that the size of each block was likely too small to lead to meaningful inference. The results turn out to be similar in the sense none of the blocks depict a significant treatment effect $\beta - \alpha$ as seen from the asymptotic 95\% confidence intervals in Table \ref{tab:analysis_MLE}.

\begin{table}
\centering
\caption{Block-wise MLEs of model parameters} \label{tab:analysis_MLE} 
\begin{tabular}{c|c|ccc|c}
 Race of & Sample size & \multicolumn{3}{c}{Parameters} & 95\% CI \\
 receiver  & $n$  & $\alpha$ & $\beta$ & $\gamma$ & for $\widehat{\beta} - \widehat{\alpha}$ \\ \hline
	         &         &             &            &               &   \\ 
Black      &     393    &  0.5312  & 0.1454  &  0.2541   &   [-8.8426, 9.6142] \\
             &             &  (5.4866)  &  (5.4997) &  (7.5981)   &   \\ \hline
			 &             &          &         &                   &  \\ 
White     &    506    & 0.2700   & 0.2400  &  1.9   & [-0.0509, 0.1110]    \\ 
             &             & (0.05067) & (0.0554) & (14.3980)       \\ \hline
			 &             &          &         &                   &  \\ 
Overall    &     899     & 0.3687   & 0.2118 &  0.688       & [-0.4069, 0.7206]  \\ 
             &             & (0.3309) & (0.3417)& (3.9290)      &  \\ \hline
\end{tabular}
\end{table}

\begin{figure}[ht]
\centering
\caption{Predicted and observed proportions of responses against race levels} \label{fig:predictions}
\includegraphics[scale=0.6]{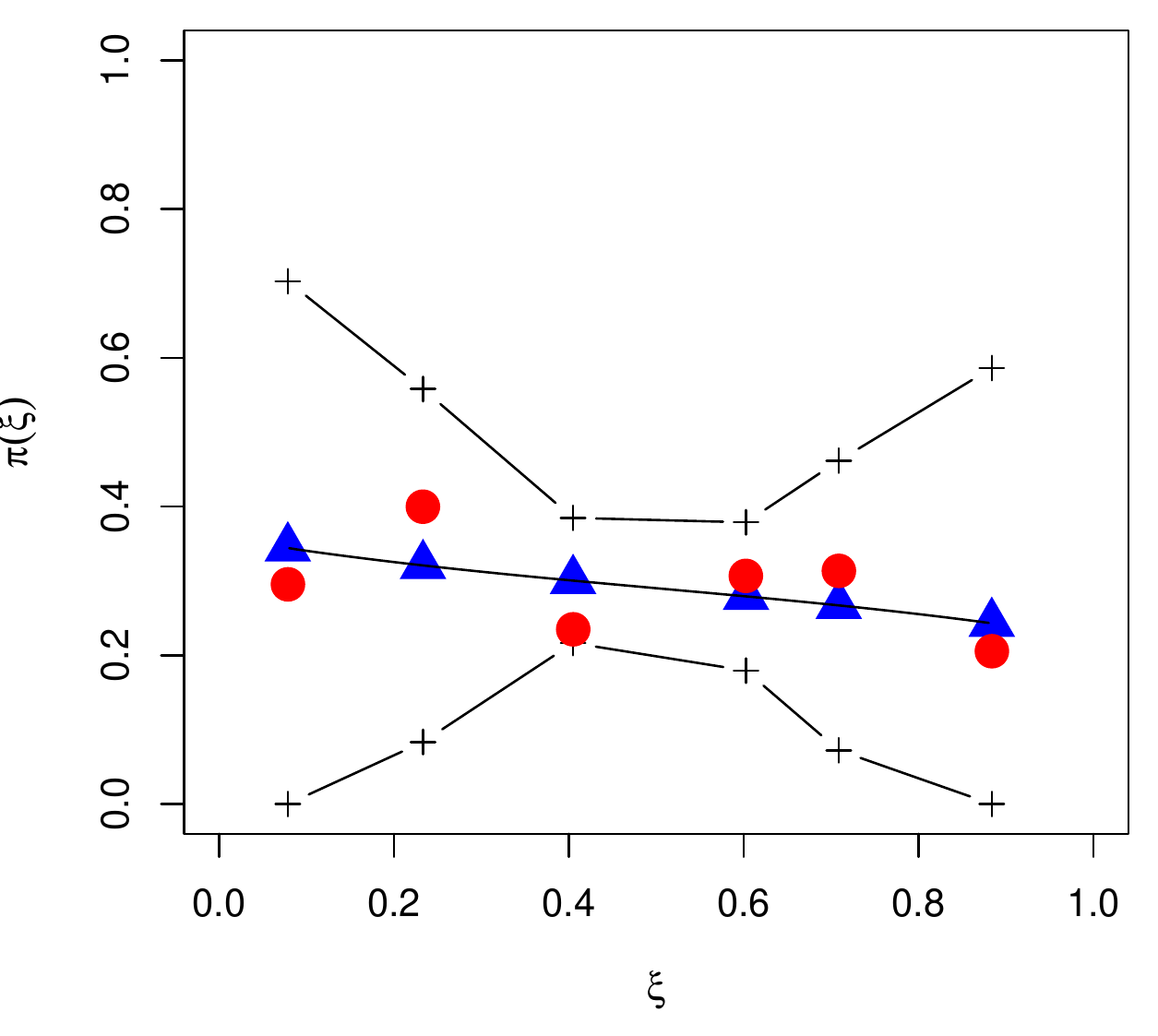} 
\end{figure}

Fig \ref{fig:predictions} shows the estimated function $\pi(\xi)$ for the experimental data. The blue dots represent estimated $\pi(\xi)$ at the applied treatment (race) levels using the fitted model, whereas the red dots represent estimated $\pi(\xi)$ at those levels using the naive proportion-based estimator. The dotted lines represent the 95\% confidence intervals for the model-based predictions. The figure shows that the best fit to the data is almost a linear function, and the wide confidence intervals at the two ends suggests that there is no evidence of any treatment effect.

\section{Discussion} \label{Sec:Discussion}

Although email experiments have been gaining in popularity in the social sciences, careful planning of such experiments from a statistical perspective is rarely used. This paper gives an example of how principles of statistical inference, sampling and design of experiments can be collectively used to plan such experiments, and analyze experimental data in a manner consistent with the planning to draw meaningful conclusions.

In the case of this particular experiment, while resource constraints prohibited us from being able to draw conclusive inferences, we were able to demonstrate all the key steps that should be followed while designing such experiments. Our simulations suggest that meaningful inference on treatment effect and shape of the curve would require very large sample sizes, whereas the data suggests that even if in our case there existed a treatment effect, it was probably too small to be captured with the available sample size.

It may thus be fair to say, the design we have developed and presented here is ready for use in contexts where it is known that discrimination exists. In future work, we expect to further identify which state bar jurisdictions suffer from problems of lawyer-side discrimination and which ones do not. For those in which there is strong evidence that discrimination exists (i.e. California), the strengths of our design for inferences about $\alpha$, $\beta$, and $\gamma$ are more likely to prove useful. Moreover, our design is generic enough that it can be used to study discrimination in markets besides law where clients seek out services via email. Rental and employment markets are two important examples.  

There are several interesting statistical research problems that can follow up this work. First, we have assumed that the input level $\xi$ is a constant and can be chosen without noise. However, in reality the ``level'' applied will only be a point estimate of an underlying ``true level''. Incorporating such uncertainty into the model will make it more realistic. Further, it is possible to construct more comprehensive models with factorial structures and covariates associated with the experimental units (lawyers) or pool information from similar groups to make sharper inference. Pooling inference from similar blocks and populations is also an interesting possibility that can be explored.

\appendix

\section{Proofs}

\subsection{Proof of Proposition \ref{prop:asym_theta}}

Successive differentiation of each element of $\nabla \ell({\bm \theta})$ in (\ref{eq:nabla_theta}) with respect to $a$, $b$ and $\gamma$ yields the following:
\begin{eqnarray*}
\frac{\partial^2 \ell}{\partial a^2} &=& - \sum_{i=1}^N \frac {1 - y_i}{ \left\{ a + g(\gamma, \xi_i) \right\}^2 } + \frac{N}{(a+b)^2}, \\
\frac{\partial^2 \ell}{\partial a \ \partial b} &=& - \frac{N}{(a+b)^2}, \\
\frac{\partial^2 \ell}{\partial a \ \partial \gamma} &=& - \sum_{i=1}^N \frac {1 - y_i}{ \left\{ a + g(\gamma, \xi_i) \right\}^2 }\ g^{\prime}(\gamma, \xi_i), \\ 
\frac{\partial^2 \ell}{\partial b^2} &=& - \sum_{i=1}^N \frac {y_i}{ \left\{ b - g(\gamma, \xi_i) \right\}^2 } + \frac{N}{(a+b)^2} \\
\frac{\partial^2 \ell}{\partial b \ \partial \gamma} &=&\sum_{i=1}^N \frac {y_i}{ \left\{ b - g(\gamma, \xi_i) \right\}^2 } \ g^{\prime}(\gamma, \xi_i) \\
\frac{\partial^2 \ell}{\partial \gamma^2} &=& \sum_{i=1}^N  g^{\prime \prime}(\gamma, \xi_i) \left\{ - \frac{y_i}{b - g(\gamma, \xi_i)} + \frac{1-y_i}{a + g(\gamma, \xi_i)} \right\} \\
&+&  \sum_{i=1}^N g^{\prime}(\gamma, \xi_i) \left\{ - \frac{y_i}{\left\{ b - g(\gamma, \xi_i) \right\}^2} \ g^{\prime}(\gamma, \xi_i)  - \frac{1 - y_i}{\left\{ a + g(\gamma, \xi_i) \right\}^2} \ g^{\prime}(\gamma, \xi_i) \right\} 
\end{eqnarray*}

Taking negative expectation of each of the second derivatives above and substituting $\mathbb{E}[Y_i] = (a+b)^{-1} \{b - g(\gamma, \xi_i) \}$, $\mathbb{E}[1 - Y_i] = (a+b)^{-1} \{a + g(\gamma, \xi_i) \}$, after some algebra, we arrive at the expression of the Fisher information matrix as stated in Proposition \ref{prop:asym_theta}.

\subsection{Proof of Corollaries \ref{cor:alphabeta_var} and \ref{cor:pi_var}}

Corollary \ref{cor:alphabeta_var} follows after a straightforward application of the multivariate Delta theorem, noting that
\begin{eqnarray*}
\mathbf{H} = \left( \begin{array}{cc} \partial \alpha/ \partial a &  \partial \beta/ \partial a \\    
\partial \alpha/ \partial b & \partial \beta/ \partial b \end{array} \right) 
= \left( \begin{array}{cc} -b/(a+b)^2 &  -(b-1)/(a+b)^2 \\    
a/(a+b)^2 & (a+1)/(a+b)^2
\end{array} \right)
\end{eqnarray*}

\bigskip

Corollary \ref{cor:pi_var} also follows from the multivariate Delta theorem, noting that by (\ref{eq:pi_function}), 
\begin{eqnarray*}
\lambda_{\new} = \nabla \pi(\theta, \xi_{\new}) = \left( \begin{array}{c} 
\partial \pi(\theta, \xi_{\new}) / \partial a \\
\partial \pi(\theta, \xi_{\new}) / \partial b \\
\partial \pi(\theta, \xi_{\new}) / \partial \gamma
\end{array} \right)
=
\left( \begin{array}{c} 
- \frac{b - g(\gamma, \xi_{\new})}{(a+b)^2} \\
\frac{a + g(\gamma, \xi_{\new}}{(a+b)^2} \\
\frac{g^{\prime}(\gamma, \xi_{\new})}{a+b}
\end{array} \right)
\end{eqnarray*}

\section{Exploring the effect of sample size on the inference} \label{sec:chooseN}

While the number of lawyers to be chosen for our experiment was actually driven by resource constraints, as discussed in Section \ref{ss:blocks}, we wanted to get some sense regarding the potential impact of the size of the experimental units ($N$) on our inference. For this purpose, additional simulations were conducted by varying $N$ for different parameter settings, keeping the number of levels $k$ fixed at six. Two criteria were considered to be most important from the point of view of exploring the impact of $N$. First, the probability $\delta_1$ that the approximate 95\% confidence interval for $\gamma$ contains the value one, which means $1-\delta_1$ can be interpreted as the power of the Wald test \citep{Wald1943} used to test the null hypothesis $\gamma = 1$ against a two-sided alternative. In other words, $1-\delta_1$ represents the power of distinguishing a linear effect of the treatment from a non-linear effect. The probability $\delta_1$ should be close to 0.95 for $\gamma=1$ (coverage of the confidence interval) and get smaller as $\gamma$ gets larger or smaller. The second criteria is the probability $\delta_2$ that the approximate 95\% confidence interval for $\alpha - \beta$ contains zero, which means $1-\delta_2$ can be interpreted as the power of the Wald test \citep{Wald1943} used to test the null hypothesis $\alpha = \beta$ against a two-sided alternative. Since the function $\pi(\xi)$ is assumed monotonic, the difference between $\alpha$ and $\beta$ is the largest possible treatment effect, and needs to be detected with high precision when it really exists. Again, this probability should be close to 95\% when $\alpha$ and $\beta$ are close, and should decrease as they are different.

\begin{figure}[h]
\centering
\caption{Plots of $\delta_1$ and $\delta_2$ against $N$ for $\alpha=.8$, $\beta=.2$} \label{fig:N_82}
\includegraphics[scale=0.47]{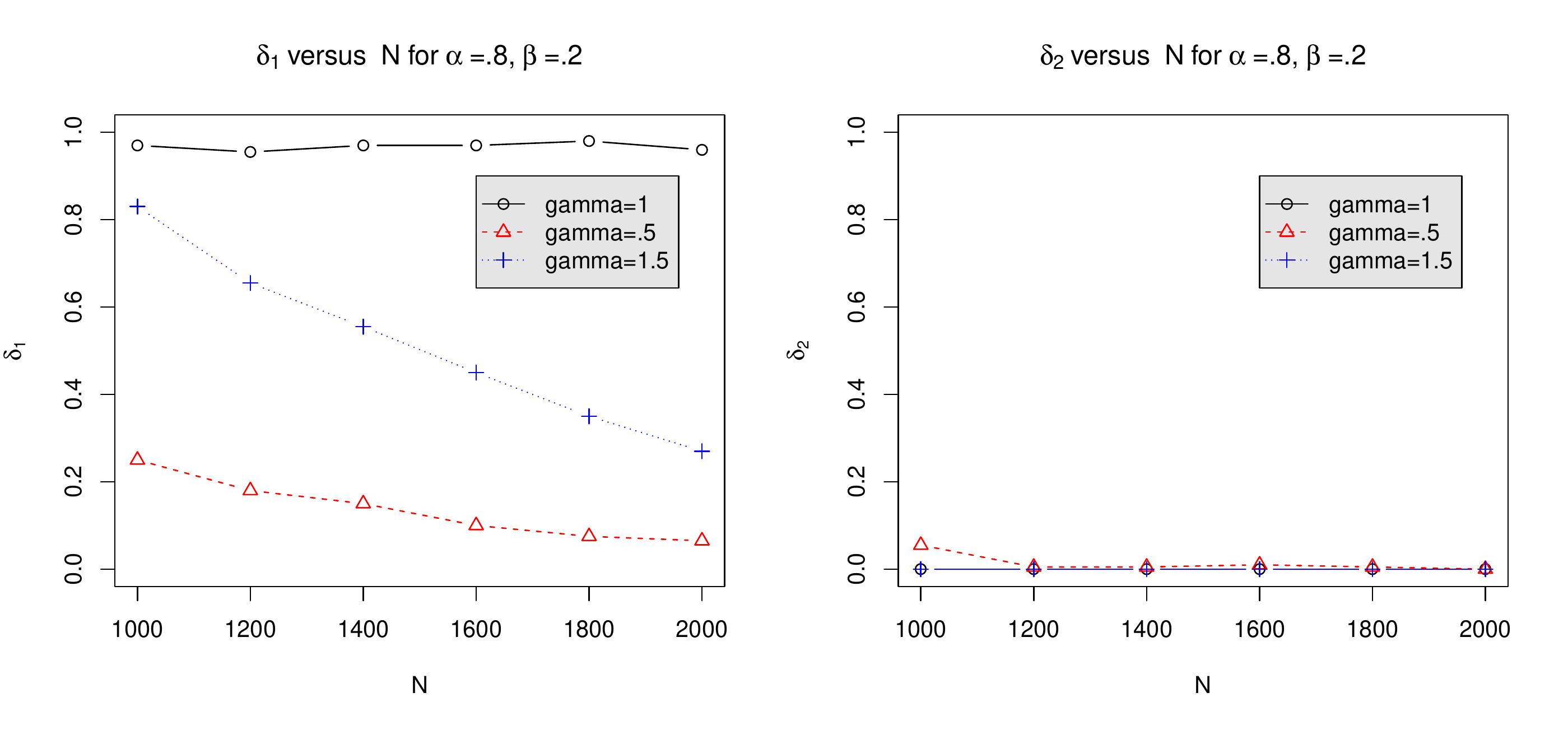}
\end{figure}

\begin{figure}[ht]
\centering
\caption{Plots of $\delta_1$ and $\delta_2$ against $N$ for $\alpha=.3$, $\beta=.2$} \label{fig:N_32}
\includegraphics[scale=0.47]{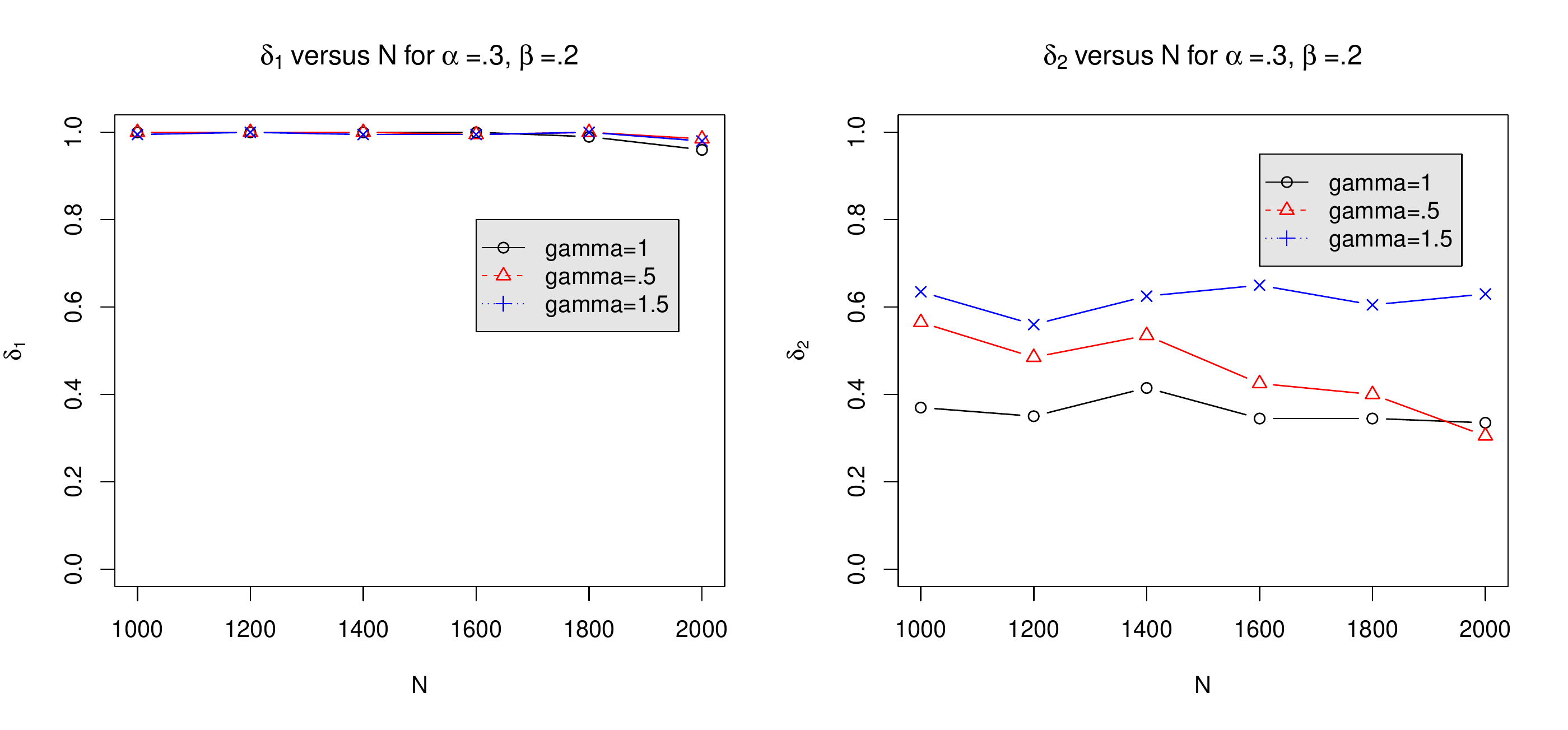}
\end{figure}

Two cases were considered in these simulations: (i) $\alpha=0.8$, $\beta=0.2$, indicating a large treatment effect, and (ii) $\alpha=0.3$, $\beta=0.2$, indicating a small treatment effect. For each of these cases, three values of $\gamma$ (0.5, 1, 1.5)  were considered. Again, for each of these six settings, the sample size $N$ was varied from 1000 to 2000 in steps of 200. Finally, for each combination of $(N, \alpha, \beta, \gamma)$, 200 datasets were generated and 95\% confidence intervals for $\gamma$ and $\alpha - \beta$ were generated. The probabilities $\delta_1$ and $\delta_2$ were estimated as the proportions of datasets that generated confidence intervals for $\gamma$ containing one and for $\alpha-\beta$ containing zero respectively. Plots of estimated $\delta_1$ and $\delta_2$ against $N$ for cases (i) and (ii) mentioned above are shown in Figures \ref{fig:N_82} and \ref{fig:N_32} respectively. 

Figure \ref{fig:N_82} shows that when there is a large treatment effect ($\alpha - \beta = 0.6$, the asymptotic intervals have approximately the true coverage rate when $\gamma = 1$ for almost all $N \ge 1000$. As $N$ increases from 1000 to 2000, the power of the Wald test to detect significant departure from linearity increases from 0.75 to 0.935 for $\gamma = 0.5$ and from 0.17 to 0.73 for $\gamma = 1.5$. Under this setting, the proposed asymptotic procedure does not have a problem detecting such a large treatment effect, irrespective of the value of $\gamma$ with $1-\delta_2$ being very close to one for all $N \ge 1000$. However, Figure \ref{fig:N_32} suggests that making inference for a small treatment effect of $\alpha - \beta = 0.1$ may be difficult even with a sample size of $N = 2000$. While the treatment effect $\alpha - \beta$ can be detected with a power varying between 0.4 and 0.6 and does not appear to reduce significantly as $N$ increases from 1000 to 2000, detection of potential non-linearity appears to be a difficult problem in such a setting. It may be remembered though, that the non-smoothness of the curves may be an artifact of the relatively small (200) number of simulations for each setting. 

Finally, one must keep in mind that the power calculations discussed above are meaningful provided the $\pi$ function exhibits a similar behavior across the experimental units. If there are blocking factors with significant effects, especially interactions with treatments, then larger sample sizes will be required to achieve the desired precision.

\bibliographystyle{chicago}
\bibliography{Lawyer_references, Gaddis}

\end{document}